\documentclass[12pt]{iopart}
\usepackage{iopams}
\usepackage{amssymb}
\usepackage{amstext}
\usepackage[english]{babel}
\usepackage[latin1]{inputenc}   
\usepackage[T1]{fontenc}        
\usepackage{setspace}
\usepackage{float}
\usepackage{longtable}
\usepackage[dvips]{color}
\usepackage{graphicx}
\usepackage{multirow}
\usepackage{subfigure}
\usepackage{xspace}
\usepackage{rotating}
\usepackage{mathrsfs}
\usepackage{cleveref}
\usepackage{upgreek}
\usepackage{tcolorbox}
\begin{document}

\title[Hydrogen localization under thermal gradients]{Hydrogen localization under thermal gradients in hydride forming metals}

\author{Karl A. Forsberg and A. R. Massih\footnote {Corresponding author: \texttt{alma@quantumtech.se}}}
\address{Quantum Technologies AB, Uppsala Science Park, Uppsala SE-75183, Sweden}

\begin{abstract}
Migration of hydrogen and hydride formation under thermal gradient leads to hydrogen redistribution in certain metals. These metals include zirconium, titanium, hafnium and their alloys with tendency to form hydrides. A computational method for hydrogen localization in such metals are presented. The method utilizes the heat flux in a steady state to compute temperature distribution (as input), and hydrogen mass flux under temperature gradient to determine hydrogen distribution both in solid solution and in the hydride phase in a two-dimensional setting. Hydrogen precipitation to hydride is determined by a solid solubility relation with an exponential function of the enthalpy of mixing per a van 't Hoff relation. The enthalpy of mixing is treated here as a stochastic variable subject to thermodynamic fluctuations. Henceforth, the Einstein-Boltzmann fluctuation theory is adapted to calculate the spatial distribution of hydrogen in solid solution and in the hydride phase. Hydrogen concentration gets localized in the  colder region of the body (Soret effect). We apply the model to the case of a zirconium alloy, Zircaloy-4, which is a material for fuel cladding utilized in pressurized water reactors. Cladding continuously picks up hydrogen due to Zr oxidation during reactor service, which we take into account. Our calculated results, hydrogen concentration profiles are comparable to experimental observations reported in the literature.
\end{abstract}


%
%
%
%
%

\section{Introduction}
\label{sec:intro}
Hydrogen absorption by group 4B metals, namely, titanium, zirconium, hafnium, can lead to formation of binary metal hydrides. The presence of hydrides in structural components that are made of these metals and their alloys,  widely used in aerospace and nuclear power industries, may be amenable to the loss of integrity since hydrides are brittle \cite{chapman2017hydrogen,Coleman_2003,bloch1997kinetics,prussner2002hydrogen}. The inflow of hydrogen in these metals/alloys are driven by several forces, namely gradients in hydrogen concentration, temperature and applied stress. More specifically, hydrogen can precipitate into hydrides when reaching its so called terminal solid solubility or TSS \cite{paton1971behavior, sinha2023terminal,une2009terminal}. Moreover, there are two TSSs one associated with precipitation (TSSP) and another with dissolution (TSSD) with an involved hysteresis \cite{qian1988hysteresis}. Most of the works reported in the literature on the topic comprise researches that were carried out on zirconium alloys, which are widely used in the core of light water reactors (LWRs), primarily as materials for fuel rod cladding that are subject to temperature, concentration and stress gradients. Concerning hydrogen transport behavior in titanium alloys, we can only mention Ref. \cite{waisman1973diffusion}.

Let us first give a synopsis of the pertinent literature. Modeling hydrogen redistribution in Zr-alloys under temperature gradient  is a venerable story. The early studies on the subject  include works of Markowitz \cite{WAPD-TM-10}, Sawatzky and Vogt \cite{sawatzky1960hydrogen,sawatzky1961mathematics}, Marino \cite{marino1972numerical,marino1974hydiz}, Maki and Sato \cite{maki1975thermal}, Bilsby \cite{bilsby1977calculation}, Byrne et al. \cite{byrne1985hydrogen}, and Takagi et al. \cite{takagi1996deuterium,
takagi2002simple}.

Marino in Ref. \cite{marino1974hydiz} provided a method for the calculation of hydrogen distribution in Zr-alloy  under a general 2-dimensional (spatial) setting. He introduced a sink term in the diffusion equation for hydrogen in solid solution, i.e., treating hydride platelets as sinks that remove hydrogen from the solution at a rate governed by hydrogen supersaturation. The rate of precipitation was assumed to be proportional to the difference between concentration of  supersaturated hydrogen in solid solution  and that in phase equilibrium taken as the terminal solid solubility limit for precipitation.  The key parameter in Marino's  model is the efficacy of the precipitation rate, which was designated by the symbol  $\alpha^2$ \cite{marino1971hydrogen}. It has the unit of inverse-time, which  needs to be determined empirically. Marino's approach was used in a subsequent study by Bilsby \cite{bilsby1977calculation} to analyze the impact of $\alpha^2$ on hydrogen and hydride distribution in a thermal gradient in Zircaloy cladding tube under reactor service conditions. Later, $\alpha^2$ was measured by Kammenzind et al. \cite{kammenzind1996hydrogen} as a function of temperature (between  561 and 633 K) and then by curve fitting the data (7 data points) obtained an Arrhenius relationship for $\alpha$. In an ensuing study, Kammenzind et al. \cite{kammenzind2000long} \textit{inter alia} measured hydrogen redistribution in hydrogenated Zircaloy-4 sheet samples under stress-free and isothermal conditions subject to a concentration gradient. They applied Marino's model to describe the measured data. However, their comparison between measured vs. calculated values was not promising. Later on, with a different modeling approach \cite{jernkvist2014multi}, this experiment was reappraised and found a fair agreement with the data.

More recent numerical approaches to the issue of hydrogen redistribution subject to temperature and stress gradients in Zr-alloys comprise the works \cite{courty2014modeling,lacroix2019zirconium,veshchunov2016new,aiyeru2023finite,nantes2024modeling}. In  \cite{courty2014modeling} the authors adapted Marino's precipitation kinetics  approach \cite{marino1971hydrogen,marino1974hydiz}, and implemented that in a 3-dimensional finite element method fuel performance code to compute hydrogen distribution in Zircaloy fuel cladding. They simulated, among other things, an accumulation of zirconium hydrides, under a radial temperature gradient,  close to the colder outer edge of Zircaloy cladding tube, which is observed in highly exposed fuel rods in light water reactors (LWRs). Veshchunov et al. \cite{veshchunov2016new} used an alternate model of hydrogen redistribution in Zr-alloy claddings during waterside corrosion in a temperature gradient. This model treats the case of  heavily precipitated hydrides which is observed in high exposure fuel cladding tubes of LWRs. Actually, they used a well-known approach for the mass transfer through a two-phase region by an additional consideration to the hydrogen supersaturation in the matrix. The  formation of the two-phase zone, i.e. hydride precipitates in the metal phase, was analysed on the basis of the Mullins-Sekerka kinetic theory of a plane front instability.

 Aiyeru et al. \cite{aiyeru2023finite} implemented a hydride precipitation and dissolution  model \cite{lacroix2019zirconium}, which accounts for the hysteresis in the phase transformation, in a commercial finite element code together with subroutines that couple the mass diffusion  of hydrogen and heat transfer that lead to redistribution and hydride precipitation in gradients of  stress and temperature. They validated the code against Sawatzky's 1960 uniaxial temperature gradient experiment \cite{sawatzky1960hydrogen} and  Kammenzind et al.'s \cite{kammenzind2000long} hydrogen concentration gradient experiment with satisfactory results. Similarly, Nantes et al. \cite{nantes2024modeling} used a finite element computer code, which includes the aforementioned hydride precipitation and dissolution model \cite{courty2014modeling,lacroix2019zirconium} to evaluate hydrogen localization in Zircaloy cladding subject to temperature gradients in LWR conditions. Their study includes hydrogen accumulation in the inter-pellet region of the cladding and the impact of missing pellet surface on hydrogen redistribution.

We should also mention the long-ago  work by Asher and Trowser (1970) \cite{asher1970distribution}, who noted that the hydrogen terminal solid solubility should be described by a fluctuating line, because hydride is precipitated preferentially on certain sites and lattice planes. Our early modeling approach to this effect was attempted in \cite{Forsberg1990redistribution}\footnote{At the time we were not aware of Asher-Trowser's study. We thank Brian Cox who later brought this paper to our attention.}, which considered the heat of mixing in the terminal solubility limit relation as a stochastic variable subject to statistical fluctuations. This model was used to explain observations of hydrogen redistribution in a boiling water reactor fuel cladding localized close to pellet interfaces.

In the present paper we revisit and augment our previous work \cite{Forsberg1990redistribution} by extending some its basic attributes and applicable capability. Our model application example is hydrogen uptake and its redistribution under axial and radial temperature gradients in Zircaloy-4 (Zr-1.3Sn-0.2Fe-0.11Cr-0.13O by wt\%) fuel cladding operating in a pressurized water reactor (PWR). The absorbed hydrogen generated by zirconium oxidation can  precipitate into hydrides ZrH$_x$ which have embrittling effect \cite{wappling1997model}. Hydrogen in solid solution of Zr-alloy is referred to as $\alpha$-phase (hcp crystal structure) while the most stable hydride ZrH$_{1.6}$ is called $\delta$-phase (fcc crystal structure) in the temperature range of interest; for details of the H-Zr system and zirconium  hydrides crystal structures see e.g. \cite{zhu2010first,zhong2012thermodynamics,lumley2014thermodynamics}.

The rest of the paper is organized as follows. In Sec. \ref{sec:eqs-formal}, we set up the formalism for obtaining the fluxes of mass and heat transport in the considered system. In Sec. \ref{sec:thermo-ss}, we describe the theoretical core of our paper. We delineate the thermal equilibrium solution for the mass flux and the formalism of energy fluctuations and the associated probability distribution, called the Einstein-Boltzmann fluctuation theory in the physics literature, adapted to our applications. From the probability distribution and the solubility limit for hydrogen, we calculate the local hydrogen concentrations in the $\alpha$-phase and the  $\delta$-phase. Section \ref{sec:apply} describes the application of the theory to the problem under consideration. The numerical method used is delineated in Sec. \ref{sec:numerics}. And in Sec \ref{sec:result}, we present the results our computations concerning hydrogen uptake of cladding at a steady rate and its redistribution with enhanced concentration close to the inter-pellet region. Because the heat flux is lower at the inter-pellet region compared to the mid-pellet region, the cladding temperature is lower there than in the latter part. Two situations, one case with normal fuel pellet geometry another  a pellet with missing surface, are evaluated there. Sections \ref{sec:discuss} and \ref{sec:sum} contain a discussion and summary of our work respectively. The paper is also supplemented by five appendices.

\

\section{Thermodynamic equations}
\label{sec:eqs-formul}
In this section, the basic thermodynamic theory and equations thereof for the problem under consideration  are presented. The formalism for the theory is described in section \ref{sec:eqs-formal}, while the way to calculate hydrogen concentration under temperature gradients accounting for energy fluctuations is discussed in section \ref{sec:thermo-ss}.
\label{sec:eqs-formal}

\subsection{Formalism}
\label{sec:eqs-formal}
The presentation here is similar to that in \cite{varias2000simulation,varias2002hydride} albeit some differences. In  thermal diffusion and heat conduction in a crystalline solid, the flux of heat $\mathbf{J}_q \equiv \mathbf{J}_1$ and that of a diffusing entity (solute), $\mathbf{J}_\mu \equiv \mathbf{J}_2$ are assumed to be linearly related through thermodynamic forces as described by the Onsager phenomenological equations \cite{onsager1931reciprocal,denbigh1951thermodynamics,deGroot2013non,howard1964matter,allnatt1967thermal,asaro2008soret}
\numparts
\begin{eqnarray}
\label{eqn:thermotrans-a}
  \mathbf{J}_1 &=& L_{11} T \nabla \Big(\frac{1}{T}\Big) + L_{12} T \nabla \Big(\frac{-\mu}{T}\Big),\\
  \mathbf{J}_2 &=& L_{21} T \nabla \Big(\frac{1}{T}\Big) + L_{22} T \nabla \Big(\frac{-\mu}{T}\Big).
  \label{eqn:thermotrans-b}
\end{eqnarray}
\label{eqn:thermotrans}
\endnumparts
In these equations, the coefficients $L_{11}$ and  $L_{22}$, which connect the flows to their associating forces, are proportional to the thermal conductivity and the diffusion coefficient of the solute, respectively.  Meanwhile,  the coefficients $L_{12}$ and  $L_{21}$ recount the coupling of the heat flow and solute flow, which in the absence of external force acting on the diffusing species, obey $L_{12}=L_{21}$ (Onsager's relation). In relations \eref{eqn:thermotrans-a}-\eref{eqn:thermotrans-b}, the symbol $\nabla$ stands for the gradient operator, $T$ is the temperature  and $\mu$ is the chemical potential, both functions of space and time. In these relations, we have not included the contribution of external forces, e.g. the stress gradient, which is considered to be negligible.

If the solute concentration and temperature gradients are small, we may suppose that $\mathbf{J}_q$ and  $\mathbf{J}_\mu$ are linear functions of $\nabla T$ and $\nabla \mu$; and following \cite{Landau_Lifshitz_1959}, we recast relations \eref{eqn:thermotrans-a}-\eref{eqn:thermotrans-b} to the form
\numparts
\begin{eqnarray}
\label{eqn:thermotrans2b}
  \mathbf{J}_q &= -L_{11} \nabla T - L_{21} T \nabla\mu + \mu \mathbf{J}_\mu,\\
  \mathbf{J}_\mu &= -L_{21} \nabla T - L_{22}\nabla \mu,
\end{eqnarray}
\endnumparts
with only three independent coefficients $L_{11},L_{21},L_{22}$. We eliminate $\nabla\mu$ from the expression for the heat flux, replacing it by $ \mathbf{J}_q$ and $\nabla T$, and write
\numparts\label{eqn:thermotrans2c}
\begin{eqnarray}
\label{eqn:thermotrans2ca}
  \mathbf{J}_q &=  \Big(\mu + \frac{L_{21}}{L_{22}} T  \Big) \mathbf{J}_\mu -k_\mathrm{th} \nabla T, \\
  \mathbf{J}_\mu &= -L_{21} \nabla T - L_{22}\nabla \mu,
  \label{eqn:thermotrans2cb}
\end{eqnarray}
\endnumparts
where
\begin{equation}\label{eqn:kappa}
k_\mathrm{th} = L_{11}-\frac{L_{21}^2}{L_{22}}T.
\end{equation}
Note that if $\mathbf{J}_\mu=0$, we have pure thermal conduction, i.e. $ \mathbf{J}_q=-k_\mathrm{th} \nabla T$, where $k_\mathrm{th}$ is the thermal conductivity of the material.

We next replace the variable $\nabla\mu$ with the gradient of the solute concentration $\nabla n$ under constant pressure by writing
\begin{equation}\label{eqn:mu-to-n}
  \nabla\mu = \Big( \frac{\partial\mu}{\partial n} \Big)_T \nabla n + \Big( \frac{\partial\mu}{\partial T} \Big)_n \nabla T.
\end{equation}
Inserting this relation in Eqs. \eref{eqn:thermotrans2ca}-\eref{eqn:thermotrans2cb} and putting
\begin{eqnarray}\label{eqn:diffusivity}
  D &= L_{22} \Big( \frac{\partial\mu}{\partial n} \Big)_T, \\
   \kappa D &= L_{22} \Big( \frac{\partial\mu}{\partial T} \Big)_n T +  L_{21} T,
\end{eqnarray}
we obtain
\numparts\label{eqn:thermotrans2d}
\begin{eqnarray}
  \mathbf{J}_q &=  \Big[ \kappa \Big(\frac{\partial\mu}{\partial n}\Big)_T -T\Big(\frac{\partial\mu}{\partial T}\Big)_n  + \mu \Big] \mathbf{J}_\mu -k_\mathrm{th} \nabla T, \\
  \mathbf{J}_\mu &= -D \Big(\nabla n + \frac{\kappa}{T}\nabla T\Big).
\end{eqnarray}
\endnumparts
Here, $D$ is the solute \emph{diffusion coefficient}, $\kappa D$ is called the \emph{thermal diffusion coefficient}, and $\kappa$ is a dimensionless quantity called the \emph{thermal diffusion ratio}: $\kappa=n Q^\ast/T$, where $Q^\ast$ is the heat of transport  measured in kelvin with  the Boltzmann constant set to $k_B=1$.

We now recall the continuity equations (a.k.a. conservation laws) for solute concentration and the heat balance, respectively, in the form
\begin{eqnarray}\label{eqn:conc-conserv2}
  \frac{\partial n}{\partial t} &= -\nabla \cdot \mathbf{J}_\mu, \\
  c_p\frac{\partial T}{\partial t}  &= - \nabla \cdot \mathbf{J}_q,
  \label{eqn:heat-conserv2}
\end{eqnarray}
where $c_p$ is the specific heat at constant pressure and the fluxes are expressed as (cf. \ref{sec:balance})
\begin{eqnarray}\label{eqn:conc-flux}
\mathbf{J}_\mu &= -D \Big(\nabla n + \frac{n Q^\ast}{T^2} \nabla T\Big), \\
\mathbf{J}_q  &=  -k_\mathrm{th} \nabla T.
  \label{eqn:heat-flux}
\end{eqnarray}
We  note that Eq. \eref{eqn:conc-conserv2} should be supplied by a source term for the solute, e.g.,  hydrogen production in the specimen denoted as  $n_g$.  We will treat this as the initial condition at every time step during the production.

\subsection{The steady state solution and energy fluctuation }
\label{sec:thermo-ss}
Let us first  express the hydrogen concentration flux \eref{eqn:conc-flux} in the body in terms of the inverse temperature $\beta \equiv T^{-1}$:
\begin{equation}
\label{eqn:conc-flux-alpha}
\mathbf{J}_\mu = -D \big(\nabla n  -  Q^\ast \, \nabla \beta \; n \big).
\end{equation}
In situations where the hydrogen production rate is too slow relative to the hydrogen diffusion rate (diffusion coefficient), a steady state (thermal equilibrium) will hold sway.  A closed system held in a temperature gradient, i.e. $\nabla \beta \ne 0$, will reach thermal equilibrium specified by $\mathbf{J}_\mu=0$ \cite{allnatt1993atomic}, so we obtain
\begin{equation}
\label{eqn:conc-flux-alpha-ss}
\frac{\nabla n}{n}= Q^\ast  \, \nabla\beta.
\end{equation}
If $Q^\ast$ is independent of temperature, we can immediately integrate Eq. \eref{eqn:conc-flux-alpha-ss} to obtain hydrogen concentration in solid solution ($\alpha$-phase) $n=n_\alpha$, viz.
\begin{equation}
\label{eqn:conc-alpha-ss}
n_\alpha= \mathcal{C} e^{\beta Q^\ast},
\end{equation}
where $\mathcal{C}$ is a constant to be determined by the boundary condition. Equation \eref{eqn:conc-alpha-ss} is the Boltzmann relation defining thermal equilibrium at which the total particle flux is zero. For the system under consideration, we assume that the temperature at the outer surface is constant at $T=T_0$ or $\beta=\beta_0$, corresponding to a solute concentration $n_u$. Hence, we write Eq. \eref{eqn:conc-alpha-ss} as
\begin{equation}
\label{eqn:conc-alpha-ss-bc}
n_\alpha= n_u e^{-(\beta_0-\beta) Q^\ast}.
\end{equation}
When the concentration $n_\alpha$ reaches a threshold value $n_s$, referred to as the terminal solid solubility for precipitation, a phase separation occurs and hydrogen precipitates into the hydride $\delta$-phase (e.g. ZrH$_{1.6}$). The precipitation process comprises nucleation in supersaturated solution and subsequent nucleus growth \cite{takagi2002simple}. For hydrogen, $n_s$ is a strong function of temperature. In thermal equilibrium, it is usually expressed by the van 't Hoff relation \cite{Coleman_2003,kearns1967terminal,une2009terminal,vizcaino2015terminal}
\begin{equation}
\label{eqn:tssd}
n_s= n_c e^{-\beta H},
\end{equation}
where $n_c$ is an empirical constant and $H$ is the enthalpy or heat of mixing also determined by measurement. At the ($\alpha,\delta$) phase boundary $n_\alpha=n_s$, we can use Eq. \eref{eqn:tssd} to write the enthalpy of mixing as
\begin{equation}
\label{eqn:enthalpy-mix}
H = - \frac{1}{\beta} \ln\frac{n_\alpha}{n_c}.
\end{equation}
We regard $H$, as in \cite{Forsberg1990redistribution}, to be a stochastic variable subject to statistical thermodynamic fluctuations. In another word, for a system in contact with a reservoir at a steady state, an extensive quantity such as $H$ will not be a constant but will fluctuate around its\emph{ mean} (a.k.a. \textit{expectation}) denoted as  $\mathbb{E}(H)$, which corresponds to its equilibrium value  with the deviation
\begin{equation}
\label{eqn:enthalpy-dev}
\boldsymbol{\updelta} (H) = H - \mathbb{E}(H).
\end{equation}
The expectation value $\mathbb{E}(H) \equiv \widehat{H}$ corresponds to a single measured value obtained by the instrument. To these  enthalpy fluctuations we assign  a probability distribution (density) function for different enthalpy values about the expectation value $\mathbb{E}(H)$. One can characterize the dispersion or the spread in this probability distribution by  the standard of deviation $\boldsymbol{\upsigma}_H$, i.e.,  the root mean-square deviation from the mean:
\begin{equation}
\label{eqn:enthalpy-rmsd}
\boldsymbol{\upsigma}_H =  \sqrt{\mathbb{E} \big((H  - \mathbb{E}(H))^2\big)}.
\end{equation}
And simple algebra gives
\begin{equation}
\label{eqn:enthalpy-2ndcum}
\boldsymbol{\upsigma}_H^2 = \mathbb{E}(H^2) - [\mathbb{E}(H)]^2 \equiv \mathbb{V}(H),
\end{equation}
where $\mathbb{V}(H)$ is the  variance (a.k.a. the second cumulant of $H$) of the probability density function. It can be related to the heat capacity at constant volume $C_V$, namely $\mathbb{V}(H) = \beta^2 C_V$. This is a special case of the \emph{fluctuation-dissipation theorem} \cite{huang2008statistical}; see also  \S 2.2 (p. 28) in \cite{martin1979statistical}.

Next, we introduce a dimensionless variable according to
\begin{equation}
\label{eqn:enthalpy-dev-rat}
x = \frac{H - \mathbb{E}(H)}{\boldsymbol{\upsigma}_H} = \frac{\boldsymbol{\updelta}(H)}{\boldsymbol{\upsigma}_H},
\end{equation}
which is a measure of the fluctuation in the enthalpy. Then at the ($\alpha,\delta$) phase boundary, by substituting for $H$ from Eq. \eref{eqn:enthalpy-mix}, we find
\begin{equation}
\label{eqn:enthalpy-dev-rat2}
x= -\Big(\frac{\beta^{-1} \ln(n_\alpha/n_c) + \mathbb{E}(H)}{\boldsymbol{\upsigma}_H}\Big).
\end{equation}
The associated probability density to this variable according to classical fluctuation (Einstein-Boltzmann) theory is Gaussian, expressed by
\begin{equation}
\label{eqn:prob-density}
p(x) = (2\pi)^{-1/2} \exp(-x^2/2)\nonumber.
\end{equation}
Then the total probability distribution function $\mathcal{P}(x)$ is
\begin{equation}
\label{eqn:enthalpy-gauss}
\mathcal{P}(x) =  \int_{-\infty}^{x} p(t)dt =\frac{1}{\sqrt{2\pi}} \int_{-\infty}^{x} e^{-t^2/2} dt,
\end{equation}
where $x$ is defined by Eq. \eref{eqn:enthalpy-dev-rat2}; see \ref{sec:einstein} for a derivation and references. The integral \eref{eqn:enthalpy-gauss}  is an error integral,  yielding
\begin{equation}
\label{eqn:error-fun-sol}
\mathcal{P}(x) = \frac{1}{2} \left(\text{erf}\left(\frac{x}{\sqrt{2}}\right)+1\right),
\end{equation}
where $\text{erf}(z)$ is the Gauss error function, which is in-built in programming language platforms such as $\mathtt{MATLAB}$  and $\mathtt{Mathematica}$. We have used the \emph{polynomial and rational approximation} method \cite{Abramowitz_Stegun_1964} to compute $\mathcal{P}(x)$ and verified our result against that of $\mathtt{Mathematica}$; see \ref{sec:pdf-compute}.

Having determined $\mathcal{P}(x)$,  the total hydrogen concentration field $n$ in the system can be calculated according to \cite{Forsberg1990redistribution}
\begin{equation}
\label{eqn:tot-conc}
n = [1-\mathcal{P}(x)] n_\alpha + \mathcal{P}(x) n_\delta,
\end{equation}
where $n_\delta$ denotes hydrogen concentration in $\delta$-phase at the ($\alpha+\delta,\delta$) phase boundary  which has a constant value. In the next section, we apply the aforementioned relations to compute  redistribution of hydrogen in a hollow cylinder subject to temperature gradients.

\section{Application: Numerical computation}
\label{sec:apply}
In this section, we apply the equations and formulae described in the preceding section to the case of redistribution of hydrogen concentration in a fuel rod cladding tube operating in a PWR.\footnote{A PWR fuel rod consists of UO$_2$ cylindrical pellets surrounded by Zr-base cladding tube; see e.g. Ref. \cite{rod2004} for various designs and Ref. \cite{ding2021review} for a recent review.} We assume the cladding tube is made of Zircaloy-4, which picks up hydrogen at a steady rate during reactor service primarily by water side corrosion of zirconium; see \ref{sec:h-uptake}. The cladding tube is subject to radial and axial temperature gradients that cause redistribution of hydrogen in the specimen.

\subsection{Numerics}
\label{sec:numerics}
For application of our theory to the deliberated structure, we note that the concentrations $n_\alpha$, $n_\delta$ are local variables, i.e., they are computed at one position of the tube wall in the $(r,z)$-plane. In order to find the total concentration across the tube wall, these variables are integrated radially as
\begin{equation}
\label{eqn:rad-conc}
\tilde n(r_w,z) = \int_{0}^{r_w} n(r,z) dr,
\end{equation}
where $r_w$ is the wall thickness and $n(r,z)$ denotes the concentration at the position $(z,r)$. The integration is done numerically across tube's wall.

We should point out  that the parameter $n_u$ entering $\mathcal{P}(x)$ through the variable $x$, defined by Eq. \eref{eqn:enthalpy-dev-rat}, and $n_\alpha$, per Eq. \eref{eqn:conc-alpha-ss-bc}, is yet unknown. That is,  $n_u$ is not given \emph{a priori} by the material properties of the system under consideration. Nevertheless, the axial-averaged total concentration $\hat{n} = \langle \tilde{n}\rangle_z$ must obey the condition of conservation of the number of atoms, namely
\begin{equation}
\label{eqn:conserve-no}
\hat{n} = n_g,
\end{equation}
where $n_g$ is the hydrogen source term in the structure which is \emph{a priori} known or precalculated (see \ref{sec:h-uptake})  as an initial condition. The unknown variable implicit in $\hat n=\hat{n}(n_u)$ is found by solving Eq. \eref{eqn:conserve-no} numerically as described in  \cite{Forsberg1990redistribution};  see the Box I  for an outline of the method.

\begin{figure}[htbp]
\begin{center}
\includegraphics[width=0.60\textwidth]{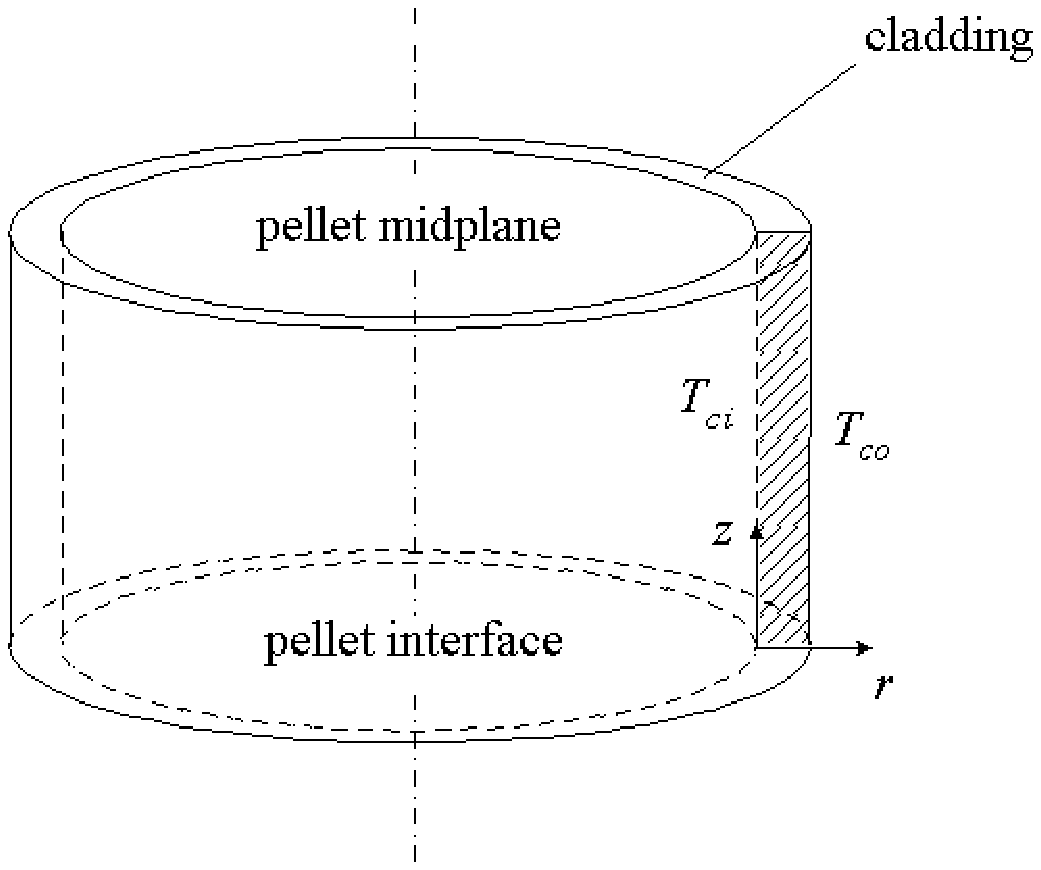}
\end{center}
\caption{Fuel pellet and cladding tube geometry schemata  where  $T_\mathrm{ci}$  and $T_\mathrm{co}$ denote cladding
 inner and outer surface temperatures, respectively.}
\label{fig:cladmod2-bmp}
\end{figure}

In order to actualize the computations, we have to prescribe an $(r,z)$-dependence for the temperature distribution in the hollow cylinder, which can be done by solving the heat equation  \eref{eqn:heat-conserv2} in a steady state with appropriate boundary conditions,  and  evaluating the ensuing radial and axial integrals numerically \cite{Forsberg1990redistribution}.\footnote{If the hollow cylinder of inner and outer radii $a$ and $b$, whose axis coincides with the axis $z$ is heated, and the boundary conditions are independent of the coordinate $\theta$, the steady state heat equation reduces to
 $\partial_r^2 T  + r^{-1} \partial_r T = 0$, with boundary condition $T(r=b,z) =T_{co}(z)$ and $k_\mathrm{th} \partial_r T|_{r=a} = - q^{\prime\prime}(z)$, where $T_{co}$ is the outer tube surface temperature  at position $z$ and $q^{\prime\prime}(z)$ is the surface heat  flux supplied at a constant rate to the inner surface of the tube at elevation $z$ of tube's axis, and $k_\mathrm{th}$ is the thermal conductivity of the tube material.} So for applying the model described in Sec. \ref{sec:thermo-ss}, we need as an input the cladding tube inner surface temperature over an axial distance around fuel inter-pellet region, where there exists a discontinuity in pellet-cladding gap due to a designed pellet chamfering, which can get magnified by pellet stacking faults and a missing pellet fragment (a.k.a. chip). Figure \ref{fig:cladmod2-bmp} shows a cartoon of a pellet surrounded by cladding. For a detailed description of a typical modern fuel pellet design see e.g. \cite{doerr2015nuclear}.

 To this end, the cladding surface temperatures at a constant heat flux corresponding to a linear heat generation rate (LHGR) of 17 kW/m and a fast neutron flux ($E \ge 1$ MeV) of $2.5 \times 10^{16}$m$^{-2}$s$^{-1}$ were calculated separately \cite{tero2004calc}. More specifically, the tube inner wall temperatures were determined for 20 axial segments over an axial distance of about 5 mm (facing the inter-pellet region)  at a cladding outside temperature of $T_{co} = 617$ K with the help of a fuel rod modeling program using the input data listed in Table \ref{tab:roddat}. And for convenience,  axial symmetry along half of the pellet region was assumed. As in  \cite{Forsberg1990redistribution}, we have found that it is sufficiently accurate to use only  the  inner surface temperature $T_{ci}$ of the tube  as input to our model  and compute the radial variation by a linear fit relation  according to
 \begin{equation}
 T(r,z)  = T_{co} + \frac{r}{r_w}\big(T_{ci}(z)-T_{co}\big),\qquad 0\le r\le r_w,
  \label{eqn:radtemp-wall}
 \end{equation}
 where as before $r_w$ denotes the wall thickness  of the tube.  The radial integration was done by Simpson's rule over a number of radial slices. We should recall that temperature drop across a hollow cylinder  wall falls logarithmic with radius, from the hotter side ($T_{ci}$) to the cooler side $T_{co}$ \cite{carslaw1959conduction}. However, for thin tubes a linear drop in temperature such as Eq. \eref{eqn:radtemp-wall} is adequate.

\begin{tcolorbox}\label{box:myboxI}
\textbf{Box I: Newton's method for solving eqn. $\hat{n}(n_u)=n_g$ for $n_u$}
\begin{itemize}
  \item Define a function
  \begin{equation}\label{eqn:funf}
    f(n_u) = \hat{n}(n_u)-n_g
  \end{equation}
  \item Use Newton's algorithm
    \begin{equation}\label{eqn:newton}
    n_u(l+1) = n_u(l) - \frac{f[n_u(l)]}{f^\prime [n_u (l)]}
  \end{equation}
  \item with
    \begin{equation}\label{eqn:fprime1}
  f^\prime [n_u (l)] = \frac{\partial f}{\partial n_u} = \Bigg\langle \int_{0}^{r_w} \frac{\partial f}{\partial n} dr \Bigg\rangle_z
  \end{equation}
   \item Use Eqs. \eref{eqn:conc-alpha-ss-bc}, \eref{eqn:enthalpy-dev-rat2}, \eref{eqn:enthalpy-gauss}, and \eref{eqn:tot-conc}
    \begin{equation}\label{eqn:fprime2}
    \frac{\partial f}{\partial n} = \Big[\frac{e^{-x^2/2}}{\sqrt{2\pi}} (\beta{\boldsymbol{\upsigma}}_H)^{-1} \big(1-\frac{n_\delta}{n_\alpha}\big)+\big(1- \mathcal{P}(x)\big)\Big]e^{(\beta-\beta_0)Q^\ast}
  \end{equation}
\end{itemize}
\end{tcolorbox}

\noindent
\begin{table}[htb]
  \caption{Nominal fuel pellet/cladding data in units of mm. The missing pellet piece  sizes, the chip height ($z_\mathrm{MC}$) and its depth ($r_\mathrm{MC}$) are defined in figure \ref{fig:pellet-mcgeo_bmp}.}
    \vspace{-2mm}
  \label{tab:roddat}
  \begin{center}
      \begin{tabular}[h]{lll}
      \hline
     Parameter                      & A     & B   \\
      \hline
     Pellet material & UO$_2$ & UO$_2$  \\
     Cladding material & Zircaloy-4 & Zircaloy-4 \\
      Pellet length                 & 9.83   & 9.83         \\
      Pellet diameter               & 8.19 &  8.19     \\
      Chamfer size; $r$-dir.        & 0.51 & 0.51      \\
      Chamfer size; $z$-dir.        & 0.13 & 0.13      \\
      Cladding outer diameter       & 9.5 & 9.5      \\
      Cladding inner diameter       & 8.36 & 8.36      \\
      Missing pellet chip size; $r_\mathrm{MC}$   & 0  & 0.635    \\
      Missing pellet chip size; $z_\mathrm{MC}$   & 0  & 1.27   \\
      \hline
      \end{tabular}
  \end{center}
\end{table}

\begin{table}[htb]
\caption{Constant model parameters for Zircaloy-4 with $\widehat{H} \equiv \mathbb{E}(H)$  \cite{une2009terminal,asher1970distribution,hong1998thermotransport}.}
\label{tab:moddat}
\centering
\begin{tabular}{lllll}
\hline
Symbol (unit)        & Definition                      & Value &  &  \\
\hline
$Q^\ast$ (K)  & Heat of transport in $\alpha$-phase &  3608  &  \\
$n_c$ (wppm)  & Constant in eq. \eref{eqn:tssd} & 32700 &  &  \\
$\widehat{H}$ (K)  & Expectation of $\delta$-phase formation enthalpy    & 3012  &  &  \\
$\boldsymbol{\upsigma}_H$ (K) & Standard deviation of $H$  & 50    &  &  \\
$n_\delta$ (wppm) & Hydrogen content of $\delta$-phase     & 16500 &  &  \\
\hline
\end{tabular}
\end{table}

\begin{figure}[htbp]
\begin{center}
\includegraphics[width=0.50\textwidth]{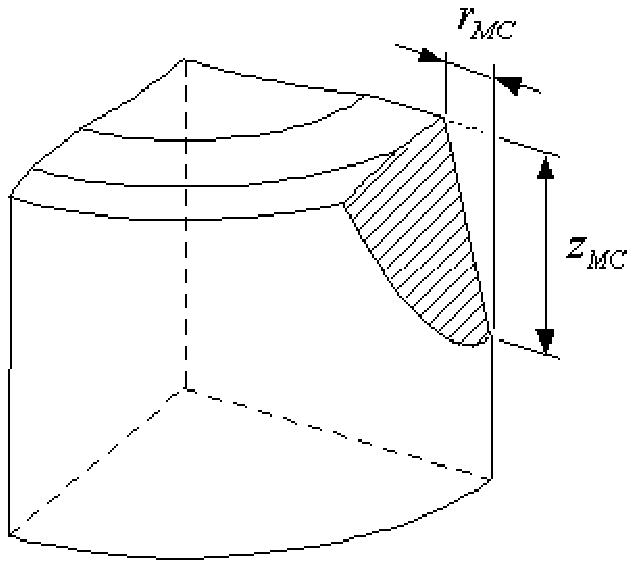}
\end{center}
\caption{An schematic geometry of  missing pellet piece (chip) defining the chip height ($z_\mathrm{MC}$) and its
 depth ($r_\mathrm{MC}$).}
\label{fig:pellet-mcgeo_bmp}
\end{figure}
\noindent
Our method has been implemented in a standalone computer program that calculates hydrogen distribution in the cladding tube opposite the inter-pellet region of the fuel rod. Numerical values for the model parameters used in our computations are given in Table   \ref{tab:moddat}.


\subsection{Results}
\label{sec:result}

\begin{figure}[htbp]
\begin{center}
\includegraphics[width=0.80\textwidth]{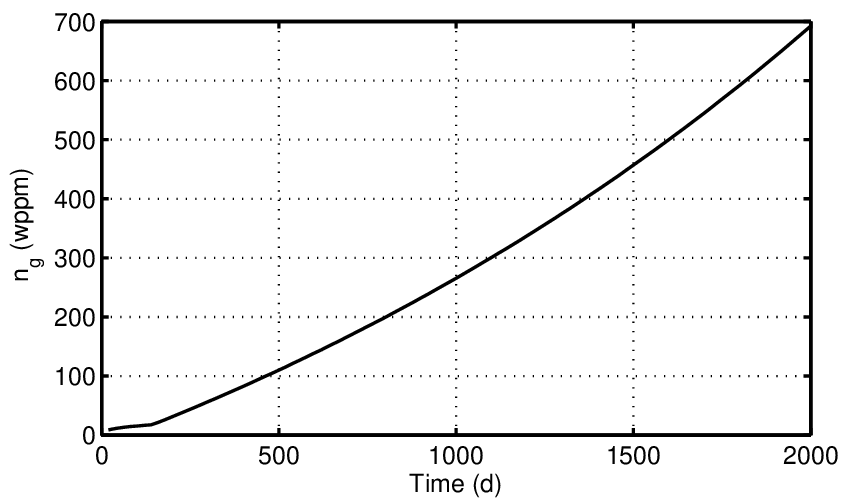}
\end{center}
\caption{Calculated total hydrogen uptake concentration in the cladding $n_g$ as a function of time; cf. \ref{sec:h-uptake}.}
\label{fig:ng-out}
\end{figure}

For our sample calculations, we consider  two cases, indicated by A and B.  Case A in Table  \ref{tab:roddat} represents the geometry of an intact pellet with inter-pellet gap due to chamfering, whereas the case B  includes also a pellet edge defect or missing chip that produces a larger pellet-cladding gap at the inter-pellet region; see  Fig. \ref{fig:pellet-mcgeo_bmp} for an idealized geometry. We carry out our computations for both cases up to 2000 days of exposure to hydrogen uptake, yielding  $n_g\approx 690$ wppm; see Fig. \ref{fig:ng-out}.   We should, however, point up at the outset that a fuel rod during irradiation in a PWR undergoes various types of deformations and changes \cite{ding2021review,van2019review}, which we do not model here. Our aim is to study the response of our  model in regard to hydrogen localization in cladding tubes subject to axial and radial temperature gradients under static conditions.

\noindent

\begin{figure}[htbp]
\centering
\subfigure[]{\label{fig:axtemp-inpa}\includegraphics[scale=0.67]{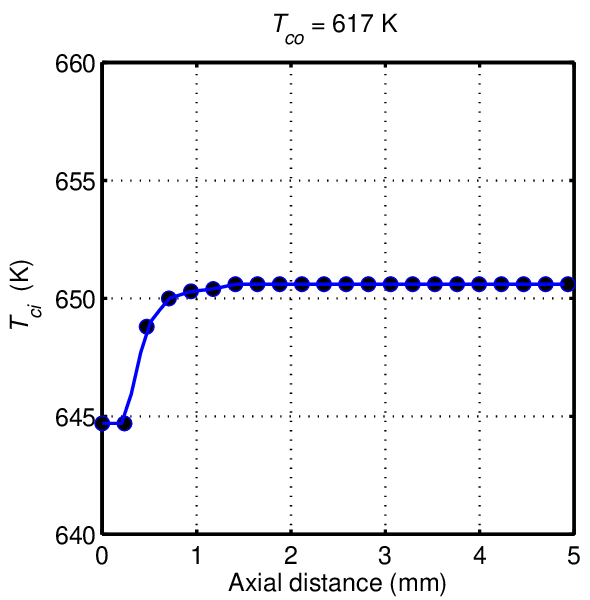}}
\subfigure[]{\label{fig:axtemp-inpb}\includegraphics[scale=0.67]{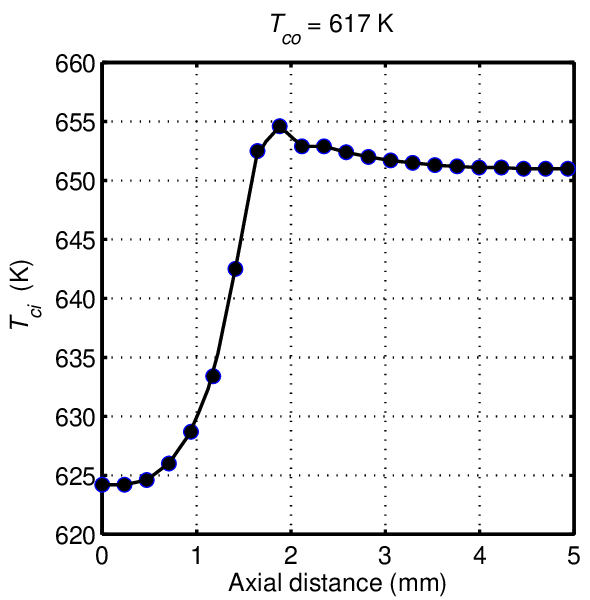}}
\vspace{1mm}
\subfigure[]{\label{fig:contour-tmpa}\includegraphics[scale=0.57]{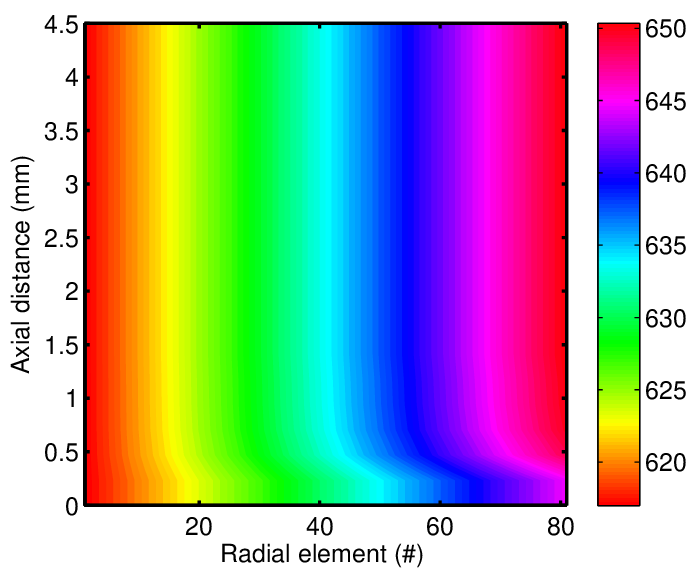}}
\hspace{0.25mm}
\subfigure[]{\label{fig:contour-tmpb}\includegraphics[scale=0.57]{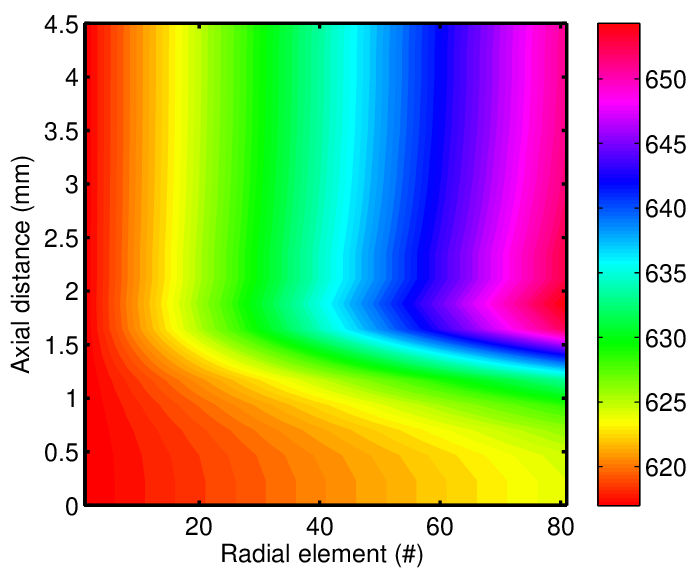}}
\caption{Cladding tube inner surface temperature $T_{ci}(z)$ close to the inter-pellet gap at $z=0$ for  case A (a), case B (b). Temperature distribution (kelvin) across cladding wall (0.57 mm) close to the inter-pellet gap for case A (c) and  case B (d). The radial element 1 is at the cladding outer radius and element 80 is located at the cladding inner radius.}
\label{fig:temp-ab}
\end{figure}

\begin{figure}[htbp]
\centering
\subfigure[]{\label{fig:h-in-alphaa}\includegraphics[scale=0.70]{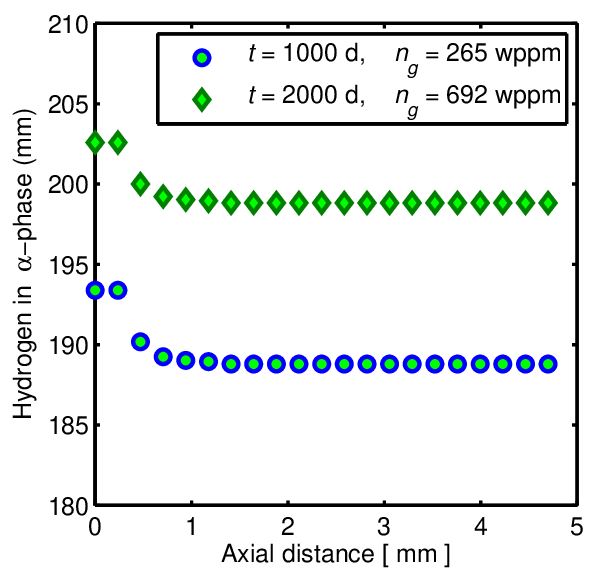}}
\subfigure[]{\label{fig:h-in-deltaa}\includegraphics[scale=0.70]{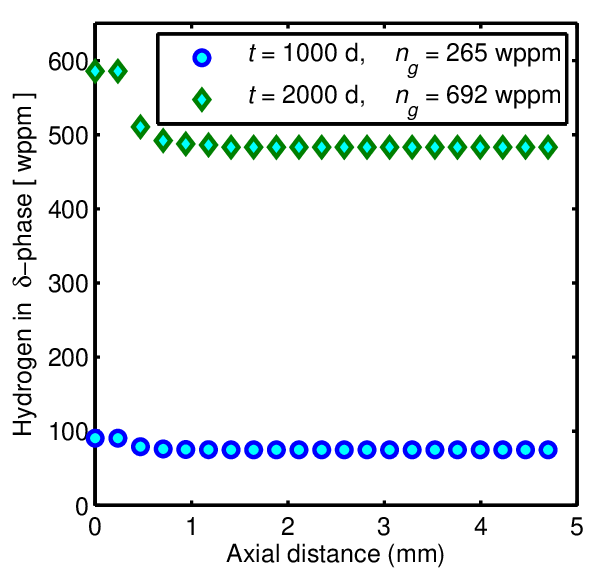}}
\caption{\textbf{Case A}: Calculated wall-averaged cladding hydrogen distribution in the vicinity of the inter-pellet gap at two instants of hydrogen uptake $n_g(t)$. (a) Hydrogen in $\alpha$-phase and (b) hydrogen in $\delta$-phase.}
\label{fig:case-a-alpha-delta}
\end{figure}

\begin{figure}[htbp]
\centering
\subfigure[]{\label{fig:h-in-alphab}\includegraphics[scale=0.70]{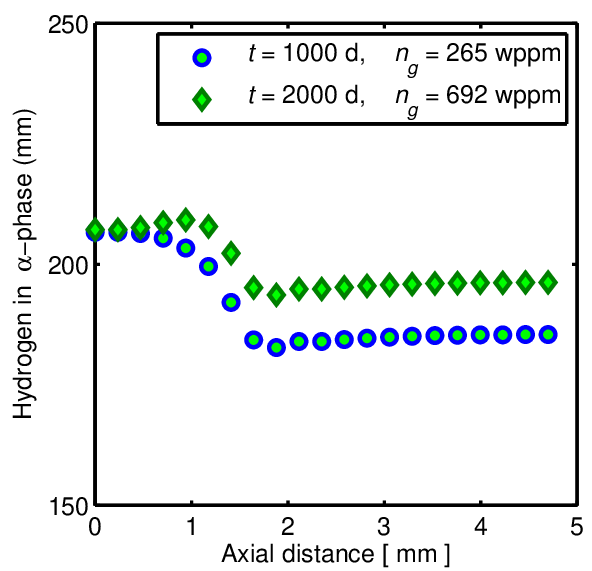}}
\subfigure[]{\label{fig:h-in-deltab}\includegraphics[scale=0.70]{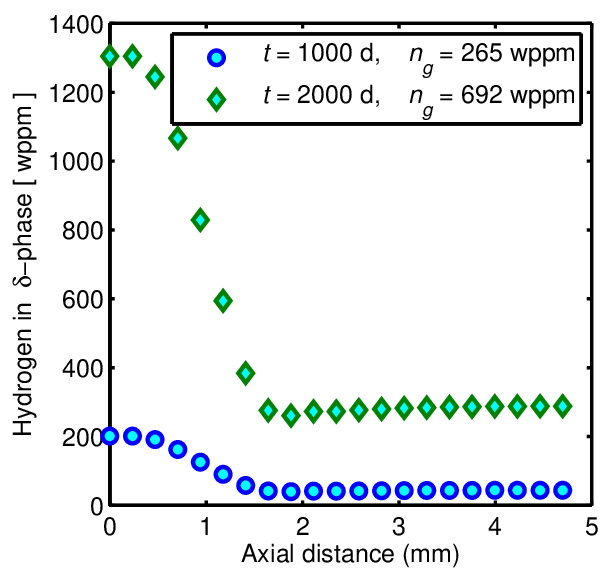}}
\caption{\textbf{Case B}: Calculated wall-averaged cladding hydrogen distribution in the vicinity of the inter-pellet gap at two instants of hydrogen uptake $n_g(t)$. (a) Hydrogen in $\alpha$-phase and (b) hydrogen in $\delta$-phase.}
\label{fig:case-b-alpha-delta}
\end{figure}

 Figures {\ref{fig:axtemp-inpa}-{\ref{fig:axtemp-inpb} depict the input  $T_{ci}(z)$ to our computations and Figs. \ref{fig:contour-tmpa}-\ref{fig:contour-tmpb}  show contours of $T(r,z)$   across the tube wall as calculated through Eq. \eref{eqn:radtemp-wall} by Simpson's numerical integration over 80 radial slices for cases A and B, respectively. The results of our computations with the described model are shown in Fig. \ref{fig:case-a-alpha-delta} for case A and in Fig. \ref{fig:case-b-alpha-delta} for case B.

 We see that the model predicts a striking surge in hydrogen localization in the $\delta$-phase after 2000 d in case B as compared to case A in the proximity of the inter-pellet gap as a result of  former's stronger temperature gradient, see Fig. \ref{fig:axtemp-inpb} versus  Fig. \ref{fig:axtemp-inpa}. This behavior is in qualitative agreement with few measured data or observations reported in the open literature; see e.g. figure 1 in \cite{Forsberg1990redistribution}, figure 10 in \cite{persson1991evaluation} and figure 3 in \cite{yang1991fuel}. Additional computational data associated with Figs. \ref{fig:case-a-alpha-delta} and \ref{fig:case-b-alpha-delta} concerning axial-radial hydrogen distribution in the cladding in the vicinity of inter-pellet gap are provided in \ref{sec:rad-ax-hdist}.

\section{Some remarks}
\label{sec:discuss}

In section \ref{sec:thermo-ss}, we alluded that in conditions that the hydrogen production rate is too slow relative to the hydrogen diffusion rate (diffusion coefficient), in the medium a thermal equilibrium would prevail. Let us be a bit more precise on this issue for the medium under consideration. From the theory of thermal diffusion, we know that the speed of the thermal diffusion is related to a characteristic time $\tau$, which for hydrogen in $\alpha$-phase is given by de Groot's formula \cite{deGroot1942theorie} as $\tau_\alpha=a^2/\pi^2D_\alpha$, where $a$ is the distance over which diffusion takes place and $D_\alpha$ is hydrogen diffusion coefficient in the $\alpha$-phase.   As detailed in \cite{deGroot1942theorie}, thermal  equilibrium is approached exponentially with time as $\propto \exp(-t/\tau_\alpha)$. Hence, when $t>\tau_\alpha$, this occurs rapidly. We dub $\tau_\alpha$ the de Groot time.

Assuming a standard form for hydrogen diffusion coefficient in Zircaloy-4 \cite{kearns1972diffusion}, $D_\alpha=D_0\exp(-E_\alpha/T)$, where $D_0=7.90\times10^{-7}$ m$^2$ s$^{-1}$ and $E_\alpha=5400$ K is the energy barrier constant,   we can estimate $\tau_\alpha$. The de Groot time at  $T=620$ K, over a mean distance of 2 mm, gives  $\tau_\alpha = 52$ minutes. This time is much less than the rate of hydrogen uptake during reactor operation, which in our case, on  average, is $\langle\dot{n}_g\rangle \approx 0.35$ wppm per day or 0.015 wppm per hour (cf. Fig. \ref{fig:ng-out}). Therefore, after a couple of hours, hydrogen reaches a state of thermal equilibrium in the $\alpha$-phase.

We may also surmise a characteristic hydrogen source time by defining the ratio  $\tau_s \equiv n_s(T)/\langle\dot{n}_g\rangle$, where $n_s(T)$ is the solubility limit for hydrogen in $\alpha$-phase at temperature $T$, which we defined earlier. Now at $T=620$ K, $n_s=254$ wppm, thereby suggesting  $\tau_s \gg \tau_\alpha$. So for the problem under deliberation, i.e. hydrogen redistribution under temperature gradients, hydrogen diffusivity does not play any role. It is the thermodynamic fluctuations that render hydrogen localization in the cladding.

From Fig. \ref{fig:temp-ab}, we see that as we move away from the inter-pellet gap and/or the pellet chip region, the cladding inside temperature levels off to its nominal value and so does the hydrogen distribution to its average value, Figs. \ref{fig:case-a-alpha-delta}-\ref{fig:case-b-alpha-delta}.  Recall that  the pellet length is about 9.8 mm, so the temperature profiles are roughly over $1/2$ the pellet length. In a recent  paper \cite{jernkvist2025hydrogen}, which has studied the same phenomenon in a boiling water reactor fuel rod (different boundary condition), a similar behavior is observed. There the temperature profile for the cladding is over 1 mm from the inter-pellet gap; see figures 1 \& 2 in \cite{jernkvist2025hydrogen}. Also, the cases considered in our previous paper with an earlier model version \cite{Forsberg1990redistribution}, the cladding  temperature profiles over 2.5 mm for pellet length of 11 mm exhibited same trend.

 Another point worth discussing is the influence of the variation in the heat of transport $Q^\ast$ for the issue under our study. Recall that $Q^\ast$ appears in the relation \eref{eqn:conc-alpha-ss-bc} for hydrogen in solid solution or $n_\alpha$ when the system is in thermal equilibrium. Because in our system  $\beta_0 \ge \beta$ ($T \ge T_0$),  $n_\alpha$ is a decreasing function of $Q^\ast$. For $T=T_0$,  $n_\alpha=n_u$, which is the same as if  $Q^\ast=0$, i.e. no heat gets transported.

 The heat of transport of hydrogen in zirconium alloys has been measured by several workers in the past \cite{sawatzky1960hydrogen,sawatzky1963heat,kammenzind1996hydrogen,hong1998thermotransport,morozumi1969effects}. Kang et al. \cite{kang2023determination} in a systematic  recent study experimentally determined $Q^\ast$ in the $\alpha$-phase of Zircaloy-4 subject to several situations. More specifically, Kang et al.'s specimens contained different hydrogen concentrations and were subjected to different uniaxial temperature gradients, representing a range of conditions. Initially, each sample had a uniform hydrogen composition, which was exposed to a linear temperature gradient. After certain durations, the  resulting hydrogen distribution was analyzed from which $Q^\ast$ was determined.

 The results of the aforementioned studies indicate that  in Zircaloy $Q^\ast$ lies in the range of $Q^\ast \approx 19-34$ kJ/mol (2285 to 4135 K). Kang et al. suggest $Q^\ast = 30\pm 5$ kJ/mol  \cite{kang2023determination}.  In our computations,  we have used  $Q^\ast =30$ kJ/mol (3608 K). We have explored a range of  $Q^\ast = 3600 \pm 600$ K with our model and found that it does not have an important impact on the outcome of our calculations, viz. the data depicted in Figs. \ref{fig:case-a-alpha-delta} and \ref{fig:case-b-alpha-delta}. However, if we increase the value of $Q^\ast$ by 50\% some effects are observed; see Table \ref{tab:qstar-effect}. We should  allude to a study \cite{jernkvist2014multi} where the authors examined a basic experiment by Sawatzky \cite{sawatzky1960hydrogen} and found that in increasing $Q^\ast$  by 40\% from its measured value would somewhat improve the outcome of their model calculations vs. experimental data; see figure 11 in \cite{jernkvist2014multi}.

\begin{table}[htp]
  \caption{Calculated effect of the heat of transport $Q^\ast$ on redistributed hydrogen concentrations in Zircaloy-4 cladding  after 2000 d of exposure at its peak/base values.}
  \label{tab:qstar-effect}
\centering
      \begin{tabular}[h]{ll|l|l}
      \hline
       Case A & \multicolumn{3}{c}{}\\
      \hline
     $Q^\ast$ (K)  & $\Longrightarrow$     & 3608    &  5412   \\
      \hline
     Min/Max $\alpha$-phase H & (wppm)     &  198/203 & 187/193 \\
     Min/Max  $\delta$-phase H  & (wppm)   &  483/586 & 494/599 \\
     \hline
      Case B & \multicolumn{3}{c}{}\\
      \hline
      Min/Max $\alpha$-phase H & (wppm)     &  196/207 & 184/205 \\
      Min/Max  $\delta$-phase H  & (wppm)   &  288/1305 & 292/1345 \\
     \hline
      \end{tabular}
\end{table}

 We should also note that for the deviation value of the enthalpy of mixing of hydrogen in $\alpha$-phase, as in \cite{Forsberg1990redistribution}, we selected  $\boldsymbol{\upsigma}_H=50$ K in our computations (Table \ref{tab:moddat}),  because it gave plausible results in \cite{Forsberg1990redistribution} versus observed data \cite{persson1991evaluation}. Varying this entity from its "nominal" value, say by up to 20\%, would not make an appreciable impact on our hydrogen redistribution results. However, increasing it by 100\% some changes on the results occur as outlined in Table  \ref{tab:varh-effect} pertaining to peak and base values  for the two cases considered here.

\begin{table}[htp]
  \caption{Calculated effect of varying $\boldsymbol{\upsigma}_H$ from its nominal value 50 K on redistributed hydrogen concentrations in Zircaloy-4 cladding after 2000 d of exposure.}
  \label{tab:varh-effect}
\centering
      \begin{tabular}[h]{ll|l|l}
      \hline
       Case A & \multicolumn{3}{c}{}\\
      \hline
     $\boldsymbol{\upsigma}_H$  (K)  & $\Longrightarrow$     & 50    &  100   \\
      \hline
     Min/Max $\alpha$-phase H & (wppm)     &  198/203 & 180/183 \\
     Min/Max  $\delta$-phase H  & (wppm)   &  483/586 & 502/608 \\
     \hline
      Case B & \multicolumn{3}{c}{}\\
      \hline
      Min/Max $\alpha$-phase H & (wppm)     &  196/207 & 175/186 \\
      Min/Max  $\delta$-phase H  & (wppm)   &  288/1305 & 319/1241 \\
     \hline
      \end{tabular}
\end{table}

Finally, we should mention that our results are in line with a recent computational study made on a PWR Zircaloy-4 rod by Nantes et al. \cite{nantes2024modeling}. Using a different model, but somewhat alike thermal boundary conditions as in our study, they calculated an axial peak hydrogen concentration of about 1150 wppm at an average hydrogen take up of 474 wppm after about 1500 days of reactor operation. The corresponding values calculated here are 790 wppm (case A) and 1510 wppm (case B) at an average hydrogen take up of 690 wppm after  2000 days of reactor operation.

\section{Summary}
\label{sec:sum}
In this paper, we first outlined the formalism of mass (hydrogen) transport in crystalline solids as described by the Onsager phenomenological equations, from which relationships for the hydrogen flux under temperature gradient and the heat flux that are bases for our computations were derived. Then, we argued that in case where the solute source term, i.e. the hydrogen production rate, is slow as compared to the characteristic time for diffusion, thermal equilibrium prevails thereby the hydrogen concentration flux $\mathbf{J}_\mu \approx 0$. This led to a Boltzmann temperature distribution for hydrogen concentration in $\alpha$-phase, viz. Eq. \eref{eqn:conc-alpha-ss-bc}. The unknown integration constant in this equation, $n_u$, was determined from the mass (concentration) balance law.

 Because in our system (Zircaloy-4 tube) hydrogen can exist in two distinct phases, namely in solid solution $\alpha$-phase and in hydride $\delta$-phase, a phase transformation could occur from $\alpha$- to $\delta$-phase when hydrogen reaches its solubility limit. The solubility limit is an exponential function of the enthalpy of mixing scaled down by temperature per van 't Hoff formula. The enthalpy of mixing was taken as a stochastic variable subject to fluctuations. We defined a dimensionless variable $x$ as a measure of the fluctuation in enthalpy. Thereafter, the probability distribution for $x$,  viz. $\mathcal{P}(x)$, was taken to be Gaussian akin to the Einstein-Boltzmann fluctuation theory.   $\mathcal{P}(x)$ was used to find the fractions of hydrogen atoms residing in the two phases.

We applied the model to the case of Zircaloy fuel cladding of a PWR, which is subject to a continuous hydrogen pickup under temperature gradients during reactor operation. For two sample biaxial temperature distributions in the cladding, denoted by A and B, we computed hydrogen distributions in the two phases of the zirconium alloy as a function of time. The results of hydrogen axial distribution were depicted in two figures (Figs. \ref{fig:case-a-alpha-delta} \& \ref{fig:case-b-alpha-delta}) for two periods of hydrogen take up (1000 d \& 2000 d). In particular, for an average hydrogen take up of 690 wppm after 2000 d of exposure, we calculated axial peak, radial average, total hydrogen concentrations of 790 wppm and 1510 wppm for cases A and B, respectively.

Our method has been implemented in a standalone computer program for computation of hydrogen distribution in the cladding tube in the region where hydrogen localization can occur due to biaxial temperature gradients. An outline of the algorithm for this program (\texttt{ZIRH4}) is given in Box II. The method  is applicable to hydrogen redistribution in a Zircaloy cladding tube exposed to a continuous hydrogen uptake during service in a light water reactor milieu. It can be integrated in a fuel rod behavior code capable of modeling local temperature variation in the cladding around inter-pellet region where gas-gap variation can occur.

\begin{tcolorbox}\label{box:myboxII}
\textbf{Box II: Algorithm for the  hydrogen redistribution program}
\begin{itemize}
  \item The program includes the innermost \emph{function} \texttt{HYDTOT} in which $n_u$, $\beta$, $\beta_0$ and $\boldsymbol{\upsigma}_H$  (sec. \ref{sec:thermo-ss}) are given as input. Then calculate $n_\alpha$ per Eq. \eref{eqn:conc-alpha-ss-bc}.
  \item Compute $\mathcal{P}(x)$  with the help of Eqs. \eref{eqn:tssd}-\eref{eqn:enthalpy-mix} and \eref{eqn:enthalpy-dev-rat2}-\eref{eqn:enthalpy-gauss}.
   \item Compute the overall hydrogen concentration via Eq. \eref{eqn:tot-conc}.

   \item These produce data along their sum and its derivative with respect to $n_u$  which is calculated by Eq. \eref{eqn:fprime2}.

   \item In the \texttt{MAIN} program a value of $n_u$ is assumed. The summed hydrogen concentration and its derivative are integrated, first radially by Simpson's rule and then axially over a pellet length.

   \item Thus $\hat{n} = \langle \tilde{n}\rangle_z$ is determined, see Eq. \eref{eqn:rad-conc} \textit{et seq.}. Then in the outermost loop  $\hat{n}$ is made to coincide with the amount of hydrogen that is there, viz. $n_g$, Eq. \eref{eqn:conserve-no}. From this $n_u$ is found by the Newton-Raphson algorithm as outlined in Box I,  in four iterations.

   \item After all these steps, the whole procedure with the exemption of the Newton-Raphson method is run one more time. This time also, $[1-\mathcal{P}(x)]n_\alpha$ and  $\mathcal{P}(x) n_\delta$ are integrated radially. And the axial-localized hydrogen concentrations in $\alpha$-phase and $\delta$-phase will be the output.
\end{itemize}
\end{tcolorbox}

\section*{Acknowledgments}
We wish to thank Lars-Olof Jernkvist for help in numerics. His suggestions made a difference in our presentation.  And we thank Tero Manng{\aa}rd for input data to our computations and the images in figures \ref{fig:cladmod2-bmp} and \ref{fig:pellet-mcgeo_bmp}. A.R.M. would like to thank Sweden's \textit{ Str{\aa}ls\"{a}kerhetsmyndigheten }(SSM) for support over the years.

\appendix
\setcounter{figure}{0}    

\section{Conservation laws}
\label{sec:balance}

The balance equations for solute concentration and the entropy are expressed, respectively,  as \cite{Landau_Lifshitz_1959}
\begin{eqnarray}\label{eqn:continuity}
  \frac{\partial n}{\partial t} + \nabla\cdot\mathbf{J}_\mu  &=& 0,  \\
  T\frac{\partial s}{\partial t} + \nabla\cdot(\mathbf{J}_q-\mu\mathbf{J}_\mu) &=& 0,
\end{eqnarray}
where $s$ stands for the specific entropy. Substituting for $\mathbf{J}_q$ and $\mathbf{J}_\mu$ from equations in \eref{eqn:thermotrans2d} and transforming the derivative $\partial s/\partial t$ as\footnote{We note that in elastic solids $s(T)=s_0(T)+K\alpha \textrm{Tr}(\boldsymbol{\varepsilon})$, where $\alpha$ is the coefficient of thermal expansion, $K$ the bulk modulus, $\boldsymbol{\varepsilon}$ is the strain tensor and $s_0$ is the entropy in the undeformed state \cite{Landau_Lifshitz_1970}. Here, we have assumed  $s=s_0(T)$.}
\begin{equation}\label{eqn:s-to-t-deriv}
  \frac{\partial s}{\partial t} = \frac{c_p}{T}\frac{\partial T}{\partial t} - \Big(\frac{\partial \mu}{\partial T}\Big)_n \frac{\partial n}{\partial t},
\end{equation}
 and after a simple algebra, we obtain
\begin{eqnarray}\label{eqn:conc-conserv}
  \frac{\partial n}{\partial t} &= D \Big(\nabla^2 n + \frac{\kappa}{T} \nabla^2 T\Big), \\
  \frac{\partial T}{\partial t}  &= \frac{c_v-c_p}{\alpha c_p} \frac{\partial}{\partial t} \nabla \cdot \boldsymbol{u} +\frac{\kappa}{c_p} \Big(\frac{\partial \mu}{\partial n}\Big)_T\frac{\partial n}{\partial t} + \chi \nabla^2 T,
  \label{eqn:heat-conserv}
\end{eqnarray}
where  the first  term on the right hand side of Eq. \eref{eqn:heat-conserv} arises by allowing the solid to deform elastically \cite{Landau_Lifshitz_1970},  $c_p$ and $c_v$ are the specific heats at constant pressure and constant volume, respectively, $\boldsymbol{u}$ is the displacement vector, related to the elastic strain $\boldsymbol{\varepsilon}$ via $\nabla \cdot \boldsymbol{u}=\textrm{Tr}(\boldsymbol{\varepsilon})$, $\alpha$ is the coefficient of thermal expansion and $\chi =k_\mathrm{th}/c_p$.

The heat balance equation   \eref{eqn:heat-conserv} however, can further be  simplified for the problem under consideration. In solids the difference between the two specific heats ($c_p-c_v$) is usually very small, thereby the first term on the right-hand side of Eq. \eref{eqn:heat-conserv} can safely be neglected;  and  the second term also is considered to be small for dilute solute concentration in solid solution  relative to the third term and may be omitted. Actually, the chemical potential can be related to the solute  concentration $n$  in an ideal solid solution via\footnote{Cf. section 5 of \cite{howard1964matter} or chapter 5 in \cite{allnatt1993atomic} for a more general formula.}
\begin{equation}\label{eqn:chempot-c}
  \mu = \mu_0 +k_BT \ln(n/n_0),
\end{equation}
where $n_0$ is the equilibrium solute concentration, $\mu_0$ is the chemical potential for the concentration $n_0$. Substituting for $\kappa=n Q^\ast/T$  with $k_B=1$, as  defined in the main text, the balance equations become
 \begin{eqnarray}\label{eqn:conc-conserv-red}
  \frac{\partial n}{\partial t} &= D \Big(\nabla^2 n + \frac{n Q^\ast}{T^2} \nabla^2 T\Big), \\
  c_p\frac{\partial T}{\partial t}  &=  Q^\ast \frac{\partial n}{\partial t} +   k_\mathrm{th}   \nabla^2 T,
  \label{eqn:heat-conserv-red}
  \end{eqnarray}
where we used Eq. \eref{eqn:chempot-c}. The first term on the right-hand side of Eq. \eref{eqn:heat-conserv-red} is considered to be small relative to the second term and therefore is omitted to obtain the usual Fourier equation for heat conduction
\begin{equation}\label{eqn:fourier}
 c_p\frac{\partial T}{\partial t}  =   k_\mathrm{th}  \nabla^2 T.
\end{equation}

\section{Einstein's formula for fluctuations in a macroscopic variable}
\label{sec:einstein}

Let us consider a closed system, and let $X$ stand for some physical quantity such as the system internal energy or the enthalpy, or entropy, etc. We denote the expectation value of this quantity, determined by measurement, by $\mathbb{E}[X]$. Now according to the Boltzmann recipe, the probability of $X$ to have a  value  is related to the entropy $S(X)$ as $S(X)=k_B\log \mathcal{W}(X) + const.$, where $k_B$ is the Boltzmann constant and $\mathcal{W}(X)$ is the probability density. Einstein \cite{einstein1910theorie} inverted this formula to express the probability for $X$ to have a value in the interval $X$ to $X+dX$ as
\begin{equation}\label{eqn:einstein}
  \mathcal{W}(X)dX = \mathrm{constant}\times e^{k_B^{-1} S(X)}dX.
\end{equation}
The entropy $S$ has a maximum at $X=X_0 \equiv \mathbb{E}[X]$, which corresponds to an equilibrium state. Therefore $\partial S/\partial X\big|_{X=X_0}=0$ and  $\partial^2 S/\partial X^2\big|_{X=X_0}<0$. The Taylor expansion of $S(X)$ to second order near its equilibrium value yields
\begin{equation}\label{eqn:entropy-expand}
S(X) = S(X_0) - \frac{1}{2} |S^{\prime\prime} (X_0)|(X-X_0)^2.
\end{equation}
Substituting this in Eq. \eref{eqn:einstein} gives the probability density in the form
\begin{equation}\label{eqn:einstein-form}
 \mathcal{W}(X) dX= A  e^{-\frac{1}{2}\gamma \big(X-X_0\big)^2} dX,
\end{equation}
where $\gamma =k_B^{-1} |S^{\prime\prime}(X_0)|$ and $A$ is the normalization constant determined by the condition
\begin{equation}\label{eqn:norm-ct}
A  \int _{-\infty}^{+\infty} e^{-\frac{1}{2}\gamma \big(X-X_0\big)^2} dX =1.
\end{equation}
The integration results in $A=\sqrt{\gamma/2\pi}$; hence
\begin{equation}\label{eqn:einstein-gauss}
\mathcal{W}(X) dX = \sqrt{\frac{\gamma}{2\pi} } e^{-\frac{1}{2}\gamma \big(X-X_0\big)^2} dX.
\end{equation}
This is  a Gaussian function first derived by Einstein \cite{einstein1910theorie} for theory of Brownian motion, discussed in classical textbooks of statistical thermodynamics \cite{Landau_Lifshitz_1980,tolman1938principles,martin1979statistical,landsberg1990thermodynamics,klimontovich1986statistical}, which we have adapted in this appendix to suit our notation. For its modern avatar see \cite{bertini2015macroscopic}.

The mean-square fluctuations or the variance is
\begin{equation}\label{eqn:meansq-fluctu}
\mathbb{E}[ (X-X_0)^2 ] = \sqrt{\frac{\gamma}{2\pi} } \int _{-\infty}^{+\infty} (X-X_0)^2 e^{-\frac{1}{2}\gamma \big(X-X_0\big)^2} dX=\gamma^{-1}.
\end{equation}
Hence, we can write the Gaussian function as
\begin{equation}\label{eqn:einstein-gauss}
  \mathcal{W}(X) dX= \frac{1}{\sqrt{2\pi \sigma_X^2}} \exp\Big[-\frac{(X-X_0)^2}{2\sigma_X^2}\Big] dX,
\end{equation}
where $\sigma_X=\sqrt{\mathbb{E} [(X-X_0)^2]} = \gamma^{-1/2}$ is the root mean-square deviation from the mean or the equilibrium value $X_0=\mathbb{E} [X] $.

Specifying now $X$ as the heat content or the heat of mixing $H$ and putting $x=(H-H_0)/\sigma_H$, with $H_0=\mathbb{E} [H] $, and relating the probability density  function $\mathcal{W}(x)$ to the the cumulative probability function $\mathcal{P}(x)$  through $\mathcal{W}(x)dx =d\mathcal{P}(x)$, we obtain the formula \eref{eqn:enthalpy-gauss} of the main text.
\newpage
\section{Computation of the Gaussian probability function}
\label{sec:pdf-compute}
In this appendix,  we compare the results of the \emph{polynomial and rational approximation} method given in \S \textbf{26.2.17} of \cite{Abramowitz_Stegun_1964}  for computing $\mathcal{P}(x)$  per Eq. \eref{eqn:enthalpy-gauss} with that  obtained  by using  Eq. \eref{eqn:error-fun-sol} directly in  \texttt{Wolfram Mathematica 13.1}. Table \ref{tab:math-as64} shows this comparison and Fig. \ref{fig:fm90Px} illustrate  the function.

\begin{table}[htb]
\caption{Computations of the function  $\mathcal{P}(x)$ by two different methods}
\vspace{-3mm}
\begin{center}
\begin{tabular}{ l|c|l }
\hline
$x$ & $\mathcal{P}(x)^a$ & $\mathcal{P}(x)^b$ \\
\hline
 0. & 0.5 & 0.500000001 \\
 0.25 & 0.598706326 & 0.598706274 \\
 0.5 & 0.691462461 & 0.691462468 \\
 0.75 & 0.773372648 & 0.773372721 \\
 1. & 0.841344746 & 0.84134474 \\
 1.25 & 0.894350226 & 0.894350161 \\
 1.5 & 0.933192799 & 0.933192771 \\
 1.75 & 0.959940843 & 0.959940886 \\
 2. & 0.977249868 & 0.977249938 \\
 2.25 & 0.987775527 & 0.987775567 \\
 2.5 & 0.993790335 & 0.99379032 \\
 2.75 & 0.997020237 & 0.997020181 \\
 3. & 0.998650102 & 0.998650033 \\
 3.25 & 0.999422975 & 0.999422914 \\
 3.5 & 0.999767371 & 0.999767327 \\
 3.75 & 0.999911583 & 0.999911555 \\
 4. & 0.999968329 & 0.999968314 \\
 \hline
\end{tabular}
\end{center}
\label{tab:math-as64}
\vspace{-3mm}
\centering
\footnotesize{ $^a$  \texttt{Mathematica 13.1}, $^b$ \S\textbf{26.2.17} of \cite{Abramowitz_Stegun_1964}.}
\end{table}
\begin{figure}[htbp]
\begin{center}
\includegraphics[width=0.60\textwidth]{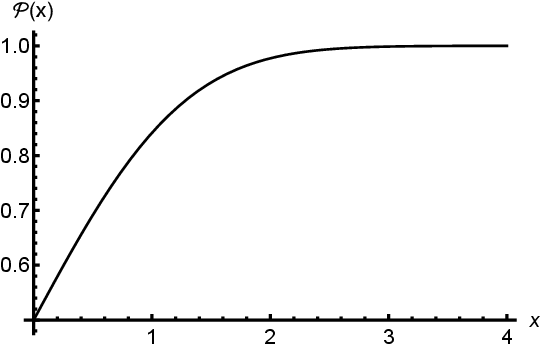}
\end{center}
\caption{The probability density function $\mathcal{P}(x)$ according to Eq. \eref{eqn:error-fun-sol}.}
\label{fig:fm90Px}
\end{figure}
\noindent
\clearpage
\section{Hydrogen production in Zircaloy tube cladding}
\label{sec:h-uptake}

The source of the hydrogen production in the cladding is the fraction of the hydrogen released by the metal-water oxidation reaction that is absorbed locally by the cladding, which is called the pickup fraction. Accordingly, our method first calculates the oxide layer growth on the cladding due to waterside corrosion, then computes the hydrogen uptake of cladding from the oxide layer thickness. For cladding oxidation we use the model described in \cite{forsberg1995model}, while for hydrogen pickup fraction we choose the method in \cite{hagrman1979matpro}.

The basic chemical reaction of the zirconium alloy with the corrosion medium water may be represented by
\[ \mathrm{Zr + 2 H_2O \to ZrO_2 + 2H_2}\]
The basic feature of the corrosion process, however,  involves three distinct regions, namely the metal, the oxide and the corrosive medium; see e.g. \cite{Northwood_Kosasih_1983}.

Zircaloy cladding oxidation during normal LWR operation is assumed to occur in two stages, which depends on the oxide layer thickness and to certain extent on the temperature of the oxide. When oxide layer is thin, oxidation rate is controlled by the entire oxide layer. However, as the oxide layer gets
thicker, a change of the outer portion occurs and further oxidation is governed by the intact inner layer \cite{hagrman1979matpro}. So, the transition between oxidation stages can be described in terms of the oxide layer thickness and temperature as is discussed below; see the venerable review report \cite{garzarolli1980review}.

The kinetics of oxidation in the first stage (pre-transition) is described by a cubic growth law according to \cite{forsberg1995model},
\begin{equation}\label{eqn:preox-rate}
 \frac{ds}{dt} = \frac{A}{s^2}\exp[-q_1/T] \quad \text{for} \quad t <t_a,
\end{equation}
and for the second stage (post-transition)  growth by a nonlinear law \cite{forsberg1995model},
\begin{equation}\label{eqn:postox-rate}
 \frac{ds}{dt} = C \big(1 + u \phi(s-s_c)\boldsymbol{\uptheta}[s-s_c]\big)\exp[-q_2/T] \quad \text{for} \quad t \ge t_a.
\end{equation}
Here,  $s$ is the oxide layer thickness (m), $ds/dt$ is the corrosion rate (m/s), $T$ is the temperature (kelvin) at the metal-to-oxide interface, roughly in the range  600 to 700 K, $t_a$ is the time to transition, $\phi$ is the fast ($E \ge 1$ MeV) neutron flux  (m$^{-2}$ s$^{-1}$) taken here as constant, $s_c$ is the threshold oxide layer thickness at which the second transition occurs, and $\boldsymbol{\uptheta}[x]$ is the Heaviside unit step function. Other parameters entering Eqs. \eref{eqn:preox-rate}--\eref{eqn:postox-rate} are constants listed in Table \ref{tab:corros-hup}. An earlier version of this model was presented in \cite{forsberg1990model}.
\begin{table}[htb]
\caption{Model parameters in Zircaloy-4 corrosion and hydrogen uptake equations. Here the oxide layer thickness unit is stated as  $\mu$m.}
\centering
\begin{tabular}{llccc }
\hline
$A$ & $1.65 \times 10^{10}$ $\mu$m$^3$ d$^{-1}$ &  & &\cite{forsberg1995model}\\
$C$ & $2.0 \times 10^{8}$ $\mu$m d$^{-1}$ &  & &\cite{forsberg1995model}\\
$q_1$ &  16250 K & & & \cite{forsberg1995model}\\
$q_2$ &  13766 K & & & \cite{forsberg1995model}\\
$q_a$ &  13150.5 K & & & \cite{forsberg1995model}\\
$t_a$ & $7.98 \times 10^{-8} \exp(q_a/T)$,  d & & & \cite{forsberg1995model}\\
$u$ & $1.70 \times 10^{-20}$  m$^2$ s $\,\mu$m$^{-1}$ & & & \cite{forsberg1995model}\\
$\phi$ & $2.5 \times 10^{16}$  $$m$^{-2}$\,s$^{-1}$ & & & \cite{geelhood2015computer}\\
$s_c$ & 6.0 $\mu$m & & &\cite{forsberg1995model}\\
$\lambda_\mathrm{ox}$ & 2.0  Wm$^{-1}$K$^{-1}$ & & &\cite{forsberg1995model}\\
$\textsf{B}$ & 0.12  & & & \cite{hagrman1979matpro}\\
\hline
\end{tabular}
\label{tab:corros-hup}
\end{table}

As noted in \cite{forsberg1995model}, in PWRs the heat flux affects the metal-oxide interface temperature. The metal-oxide temperature is a function of the heat flux, the water temperature, the conductivity of the oxide, and the oxide film thickness. The temperature change through the oxide layer film formed at a given time according to \cite{forsberg1995model} is expressed as
\begin{equation}\label{eqn:effective-temp}
 T = T_0 + \textmd{f}(s-s_0),
\end{equation}
where $s$ is the current oxide thickness (at time $t$), $s_0$, is the initial oxide thickness associated with temperature $T_0$,
which is the temperature at the metal-oxide interface at time $t_0$, and $\textmd{f}=q^{\prime\prime}/\lambda_\mathrm{ox}$, $q^{\prime\prime}$ is the heat flux (Wm$^{-2}$), and $\lambda_\mathrm{ox}$ is the oxide thermal conductivity  (Wm$^{-1}$K$^{-1}$).

In the temperature range of interest, i.e. $ 600 \lesssim T_0 \lesssim 700$ K, $\lambda_\mathrm{ox}\approx 2.0$ Wm$^{-1}$K$^{-1}$ \cite{hagrman1979matpro}, it is shown that analytical solutions to Eqs. \eref{eqn:preox-rate}-\eref{eqn:postox-rate} can be obtained \cite{forsberg1995model}. The solutions read as
\begin{equation}\label{eqn:oxrate-sol}
\label{cases}
 s(t) = \cases{\Bigg(\frac{\Sigma^3 -s_0^3 \mathfrak{c}(\Sigma-s_0)}{1-\mathfrak{c}(\Sigma-s_0)}\Bigg)^{1/3},&if  $t \leq t_a$, \\
s_0 + \frac{\exp(F G)-1}{G},& otherwise.\\}
\end{equation}
where
\begin{eqnarray}\label{eqn:Sigma}
\Sigma  &=  \Big[3(t-t_0) A \exp\big(-q_1/T_0\big) + s_0^3\Big]^{1/3},\\
\label{eqn:cfrak}
\mathfrak{c} &= \frac{q_1 \textmd{f}}{2T_0^2},\\
\label{eqn:G-capit}
G &= \frac{q_2\textmd{f}}{T_0^2} + \frac{\mathfrak{D}}{1+\mathfrak{D}(s_0-s_c)},\\
\label{eqn:F-capit}
F &= C\big[1 + \mathfrak{D}(s_0-s_c)\big]\exp(-q_2/T_0)(t-t_0),\\
\label{eqn:Dfrak}
\mathfrak{D} &= u\phi \boldsymbol{\uptheta}[s_0-s_c].
\end{eqnarray}
The derivations of the solutions in \eref{eqn:oxrate-sol} are given in appendices A and B of \cite{forsberg1995model}. Here, we have reproduced them with some minor typographical corrections.

To illustrate our result, we have plotted $s(t)$ as a function of $t$ up to $t=1000$ days at $T_0$=613 K (340 $^\circ $C) in Fig. \ref{fig:oxid-345c}. This result can be compared with the data presented in figure B-14.4 of  Hagrman and Reymann \cite{hagrman1979matpro}, which indicates fair concordance. After 1000 days of exposure, we calculate $s\approx 37\, \mu$m, while Hagrman and Reymann show $s\approx 32\, \mu$m, despite different models and computational method. The temperature dependence of the oxide layer growth as predicted by our model is illustrated in Fig. \ref{fig:oxid-multi}.
\begin{figure}[htbp]
\begin{center}
\includegraphics[width=0.75\textwidth]{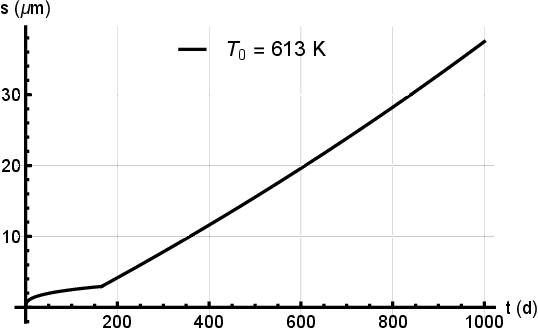}
\end{center}
\caption{Calculated oxide layer thickness $s$ on Zircaloy-4 clad vs. time $t$ (day) at 613 K (340$^\circ$C). At the transition point, $t_a=165.5$ d and $s(t_a)=2.95$ $\mu$m.}
\label{fig:oxid-345c}
\end{figure}
\noindent
\begin{figure}[htbp]
\begin{center}
\includegraphics[width=0.75\textwidth]{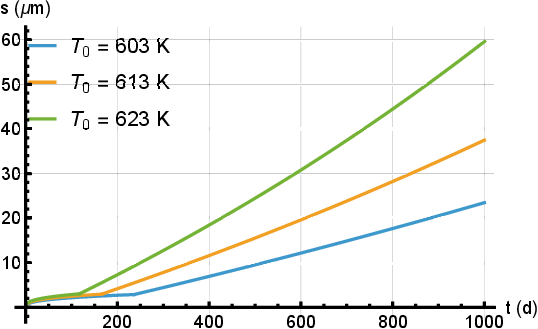}
\end{center}
\caption{Calculated oxide layer thickness $s$ on Zircaloy-4 cladding vs. time $t$ (day) at several temperatures.}
\label{fig:oxid-multi}
\end{figure}
After calculating the oxide layer thickness at the start and end of the current time step for a given metal-to-oxide temperature $T$,  we compute the average weight (mass) fraction of hydrogen uptake in the cladding at the start of the current time step. For this purpose, we utilize the procedure outlined by Hagrman  \cite{hagrman1979matpro} for the \emph{surface-controlled} hydrogen pickup fraction, which posits that the pace of hydrogen pickup with respect to oxide layer thickness (or its equivalent mass $w$) is constant, i.e. the derivative $dn_g/dw=$ const. In more detail, this quantity depends on the aforementioned oxide layer transition thickness ($s_t \propto w_t)$, namely
\begin{equation}\label{eqn:huprat}
\frac{d n_g}{dw} = \cases{
    \frac{\textsf{B}}{8},& if  $w < w_t$ \\
     \frac{\textsf{C}}{8},& otherwise.\\}
\end{equation}
Here, $n_g$ is the hydrogen mass gain per unit area ($\mu$g m$^{-2}$), $\textsf{B}$ and $\textsf{C}$ are constants, which we put $\textsf{B}=\textsf{C}=0.12$ for Zircaloy-4 in PWR  as in \cite{hagrman1979matpro}, where the number 8 accounts for different masses of hydrogen and oxygen in H$_2$O. Integrating Eq. \eref{eqn:huprat} and expressing the right-hand side of the equation in terms of the oxide layer thickness $s$, we write
\begin{equation}\label{eqn:hup-ox}
 n_g[s(t)]  = \mathfrak{f} \frac{\textsf{B}}{8} s(t),
\end{equation}
where  $\mathfrak{f}$ is a conversion factor, which makes the unit of $n_g$ in weight fraction of hydrogen (wppm). It is given by
\begin{equation}\label{eqn:convert-oxwt-hwt}
\mathfrak{f} = \frac{9 \times 10^5 d_o}{d_o^2-d_i^2},
\end{equation}
where $d_o$ and $d_i$ are the outer and inner diameter (m) of the tube, respectively.

We illustrate our result by plotting $n_g$ as a function of $t$ up to $t=1000$ days at $T_0$=613 K (340 $^\circ $C) in Fig. \ref{fig:hup-345c}.  The temperature dependence of the hydrogen pickup as predicted by the putative model is displayed in Fig. \ref{fig:hup-multi}. The Zircaloy-4 cladding tube considered here has dimensions $d_o=9.5$ mm and $d_i=8.356$ mm.

The presented results seem plausible. For example, an empirical generic correlation for Zircaloy-4 cladding in PWRs (based on post-irradiation data fit) shows that hydrogen pickup at a fuel burnup of 60 MWd(kgU)$^{-1}$ amounts to $n_g \approx 600$ wppm \cite{zhang2016limiting}. After 2000 days of irradiation, corresponding to roughly $\approx 65$ MWd(kgU)$^{-1}$, we compute with our  model and tube dimensions $n_g \approx 564$ wppm at the constant temperature of $613$ K.

\begin{figure}[htbp]
\begin{center}
\includegraphics[width=0.75\textwidth]{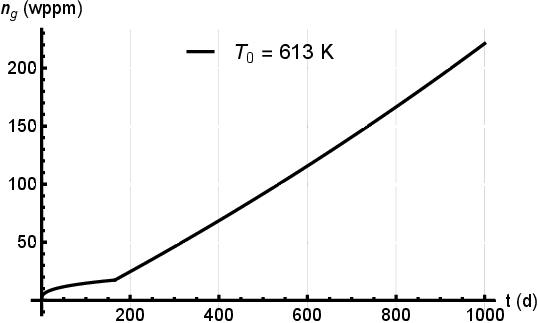}
\end{center}
\caption{Calculated hydrogen uptake $n_g$ of Zircaloy-4 clad vs. time $t$ (day) at 613 K (340$^\circ$C).}
\label{fig:hup-345c}
\end{figure}
\noindent
\begin{figure}[htbp]
\begin{center}
\includegraphics[width=0.75\textwidth]{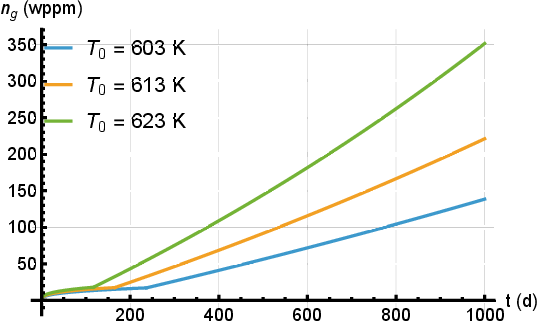}
\end{center}
\caption{Calculated hydrogen uptake $n_g$ of Zircaloy-4 cladding vs. time $t$ (day) at several temperatures.}
\label{fig:hup-multi}
\end{figure}
\clearpage
\section{Results of radial-axial hydrogen distribution in the cladding tube}
\label{sec:rad-ax-hdist}

Here, we present supplementary computational data to Figs. \ref{fig:case-a-alpha-delta} and \ref{fig:case-b-alpha-delta} of the main text, namely, the radial variation of hydrogen concentration at several axial positions in the cladding opposite to the inter-pellet region for the two considered cases A and B; see section \ref{sec:result}. The hydrogen distribution data presented below pertain to 2000 days of Zircaloy-4 cladding exposure to hydrogen uptake yielding the mean concentration of $n_g \approx 960$ wppm  in the considered cladding segment according to our model calculation.

For case A, Fig. \ref{fig:case-a-radax}(a) shows the temperature distribution across the cladding wall. The origin of the abscissa is the outer radius of the cladding tube and the tube wall is divided to 80 equidistant elements. Fig. \ref{fig:case-a-radax}(b) depicts hydrogen concentration in the $\alpha$-phase (solid solution) while Fig. \ref{fig:case-a-radax}(c) shows that in the $\delta$-phase (hydride). An interesting or peculiar behavior observed for hydrogen in solid solution in which the calculated lines go through maxima. We should mention that the displayed data correspond to predictions of Eq. \eref{eqn:tot-conc} of the main text prior to radial integration. The corresponding data for case B are displayed in  Fig. \ref{fig:case-b-radax}.

Collating Fig. \ref{fig:case-b-radax}(b) against Fig. \ref{fig:case-a-radax}(b), we notice that for case B, as one looks away from the inter-pellet region, the maxima in $\alpha$-phase hydrogen concentration gets closer to the outer wall of the cladding, which hardly occurs in case A, due to a more distorted temperature distribution in case B. Similarly, comparing Fig. \ref{fig:case-b-radax}(c) with Fig. \ref{fig:case-a-radax}(c), we see that the radial  hydrogen concentration in $\delta$-phase is much broader closer to the inter-pellet region than away from it. Moreover, the peak hydrogen concentrations at the outer wall are markedly lower in case B than in case A.

\begin{figure}[htb]
\centering
\subfigure[]{\label{fig:cladtemp-radaxa}\includegraphics[scale=0.80]{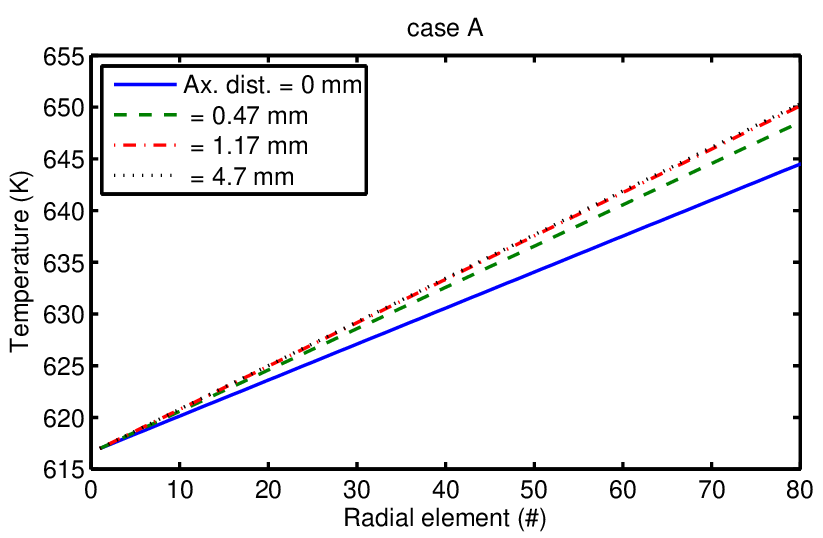}}
\subfigure[]{\label{fig:h-alpha-radaxa}\includegraphics[scale=0.67]{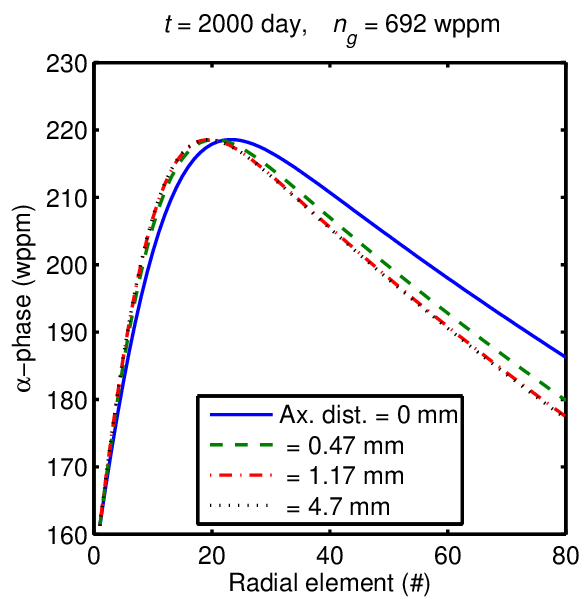}}
\subfigure[]{\label{fig:h-delta-radaxa}\includegraphics[scale=0.67]{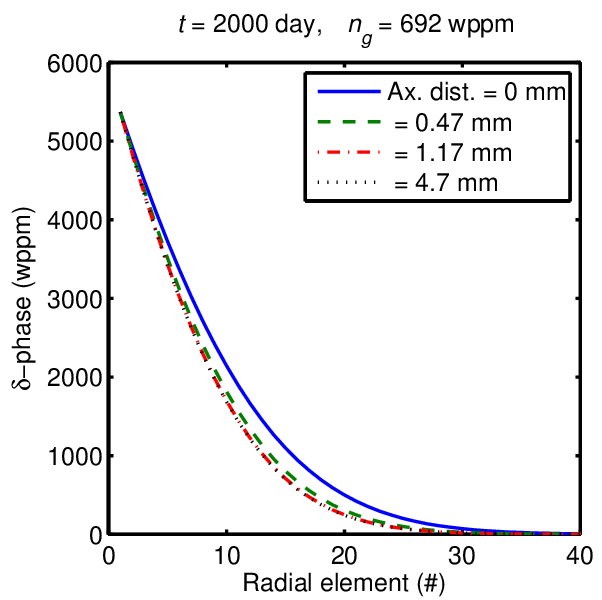}}
\caption{\textbf{Case A}: (a) Radial temperature across cladding wall (0.57 mm) at several axial distances from inter-pellet position. (b) Hydrogen concentration across cladding wall at several axial distances from inter-pellet position in $\alpha$-phase; and  (c) in $\delta$-phase. Abscissa's origin is at the outer radius of the cladding tube.}
\label{fig:case-a-radax}
\end{figure}

\begin{figure}[htb]
\centering
\subfigure[]{\label{fig:cladtemp-radaxb}\includegraphics[scale=0.80]{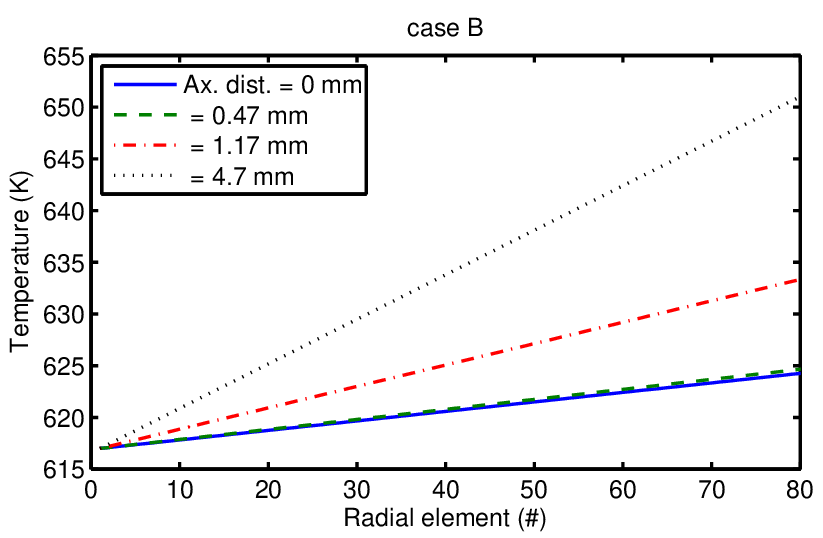}}
\subfigure[]{\label{fig:h-alpha-radaxb}\includegraphics[scale=0.67]{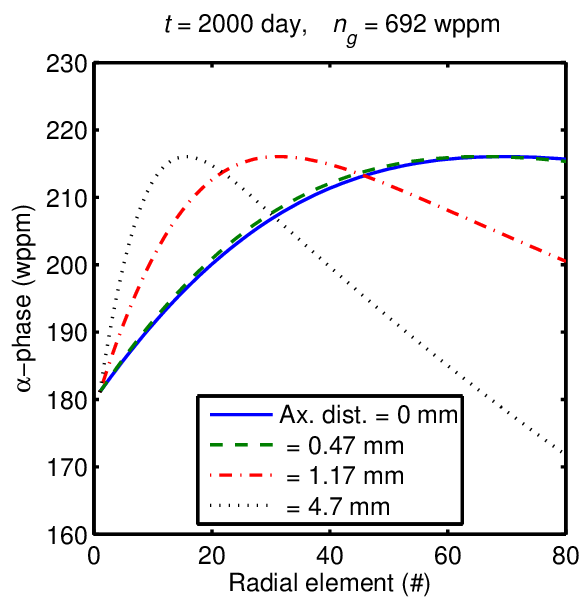}}
\subfigure[]{\label{fig:h-delta-radaxb}\includegraphics[scale=0.67]{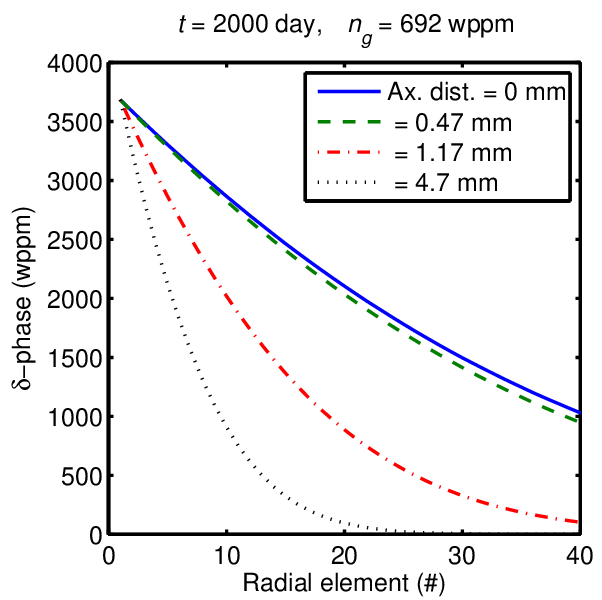}}
\caption{\textbf{Case B}: (a) Radial temperature across cladding wall (0.57 mm) at several axial distances from inter-pellet position. (b) Hydrogen concentration across cladding wall at several axial distances from inter-pellet position in $\alpha$-phase; and  (c) in $\delta$-phase. Abscissa's origin is at the outer radius of the cladding tube.}
\label{fig:case-b-radax}
\end{figure}

\clearpage

\section*{References}
\bibliographystyle{unsrt}
\bibliography{microRev2}

\end{document}